\documentclass[twocolumn]{IEEEtran}

\usepackage[mathscr]{eucal}
\usepackage[cmex10]{amsmath}
\usepackage{epsfig,epsf,psfrag}
\usepackage{amssymb,amsmath,amsthm,amsfonts,latexsym}
\usepackage{amsmath,graphicx,bm,xcolor,url}
\usepackage{fixltx2e}
\usepackage{array}
\usepackage{verbatim}
\usepackage{bm}
\usepackage{algorithmic, cite}
\usepackage{algorithm}
\usepackage{verbatim}
\usepackage{textcomp}
\usepackage{mathrsfs}
\usepackage{multirow}
\usepackage{epstopdf}
\usepackage{multicol}
\usepackage{ifpdf}
\usepackage{caption}

\catcode`~=11 \def\UrlSpecials{\do\~{\kern -.15em\lower .7ex\hbox{~}\kern .04em}} \catcode`~=13 

\allowdisplaybreaks[1]
 
\newcommand{\nn}{\nonumber}

\newcommand{\calN}{\mathcal{N}}

\newcommand{\calR}{\mathcal{R}}

\newcommand{\bH}{\mathbf{H}}

\newcommand{\bI}{\mathbf{I}}

\newcommand{\bR}{\mathbf{R}}

\newcommand{\by}{\mathbf{y}}
\newcommand{\bY}{\mathbf{Y}}

\newcommand{\bbE}{\mathbb{E}}

\newcommand{\bbP}{\mathbb{P}}

\newcommand{\bbZ}{\mathbb{Z}}

\DeclareMathAlphabet{\mathbsf}{OT1}{cmss}{bx}{n}
\DeclareMathAlphabet{\mathssf}{OT1}{cmss}{m}{sl}

\DeclareSymbolFont{bsfletters}{OT1}{cmss}{bx}{n}  
\DeclareSymbolFont{ssfletters}{OT1}{cmss}{m}{n}
\DeclareMathSymbol{\bsfGamma}{0}{bsfletters}{'000}
\DeclareMathSymbol{\ssfGamma}{0}{ssfletters}{'000}
\DeclareMathSymbol{\bsfDelta}{0}{bsfletters}{'001}
\DeclareMathSymbol{\ssfDelta}{0}{ssfletters}{'001}
\DeclareMathSymbol{\bsfTheta}{0}{bsfletters}{'002}
\DeclareMathSymbol{\ssfTheta}{0}{ssfletters}{'002}
\DeclareMathSymbol{\bsfLambda}{0}{bsfletters}{'003}
\DeclareMathSymbol{\ssfLambda}{0}{ssfletters}{'003}
\DeclareMathSymbol{\bsfXi}{0}{bsfletters}{'004}
\DeclareMathSymbol{\ssfXi}{0}{ssfletters}{'004}
\DeclareMathSymbol{\bsfPi}{0}{bsfletters}{'005}
\DeclareMathSymbol{\ssfPi}{0}{ssfletters}{'005}
\DeclareMathSymbol{\bsfSigma}{0}{bsfletters}{'006}
\DeclareMathSymbol{\ssfSigma}{0}{ssfletters}{'006}
\DeclareMathSymbol{\bsfUpsilon}{0}{bsfletters}{'007}
\DeclareMathSymbol{\ssfUpsilon}{0}{ssfletters}{'007}
\DeclareMathSymbol{\bsfPhi}{0}{bsfletters}{'010}
\DeclareMathSymbol{\ssfPhi}{0}{ssfletters}{'010}
\DeclareMathSymbol{\bsfPsi}{0}{bsfletters}{'011}
\DeclareMathSymbol{\ssfPsi}{0}{ssfletters}{'011}
\DeclareMathSymbol{\bsfOmega}{0}{bsfletters}{'012}
\DeclareMathSymbol{\ssfOmega}{0}{ssfletters}{'012}

\newcommand{\eps}{\varepsilon}

\DeclareMathOperator{\diag}{diag}

\newtheorem{theorem}{Theorem} 
\newtheorem{lemma}{Lemma}

\newtheorem{proposition}{Proposition}
\newtheorem{corollary}{Corollary}

\newtheorem{remark}{Remark}

\newcommand{\qednew}{\nobreak \ifvmode \relax \else
      \ifdim\lastskip<1.5em \hskip-\lastskip
      \hskip1.5em plus0em minus0.5em \fi \nobreak
      \vrule height0.75em width0.5em depth0.25em\fi}

\usepackage{xspace}
\usepackage[ colorlinks = true,
             linkcolor = blue,
             urlcolor  = blue,
             citecolor = red,
             anchorcolor = green,
]{hyperref}

\newcommand{\red}[1]{\textcolor{black}{#1}} 
\newcommand{\Red}[1]{\textcolor{black}{#1}} 

\usepackage{cite}
\allowdisplaybreaks[2]
\begin{document} 

\title{On the Capacity of Symmetric M-user Gaussian Interference Channels with Feedback} 

\author{Lan V.\ Truong, {\em Member, IEEE},  $\quad$ 
        Hirosuke Yamamoto, {\em Life Fellow, IEEE}  \thanks{This work was supported in part by JSPS Grant-in-Aid for Scientific Research, No.~25289111. This paper was presented in part at the 2015 International Symposium on Information Theory.}
\thanks{L.\ Truong is with the Department of Computer Science, National University of Singapore, Singapore (e-mail: truongvl@comp.nus.edu.sg).} \thanks{H.\ Yamamoto is with The University of Tokyo, Japan (e-mail: Hirosuke@ieee.org).}  \thanks{Communicated by M.\ Costa, Associate Editor for Shannon Theory.}
 }

\maketitle
\begin{abstract} A general time-varying feedback coding scheme is proposed for $M$-user fully connected symmetric Gaussian interference channels. Based on the analysis of the general coding scheme, we prove a theorem which gives a criterion for designing good time-varying feedback codes for Gaussian interference channels. The proposed scheme improves the Suh-Tse and Kramer inner bounds of the channel capacity for the cases of weak and not very strong interference when $M=2$. This capacity improvement is more significant when the signal-to-noise ratio (SNR) is not very high. In addition, our coding scheme can be proved mathematically and numerically to outperform the Kramer code for $M\geq 2$ when the SNR is equal to the interference-to-noise ratio (INR). Besides, the generalized degrees-of-freedom (GDoF) of our proposed coding scheme can be proved to be optimal in the all network situations (very weak, weak, strong, very strong) for any $M$. The numerical results show that our coding scheme can attain better performance than the Suh-Tse coding scheme for $M=2$ or the Mohajer-Tandon-Poor lattice coding scheme for $M>2$. Furthermore, the simplicity of the encoding/decoding algorithms is another strong point of our proposed coding scheme compared with the Suh-Tse coding scheme when $M=2$ and the Mohajer-Tandon-Poor lattice coding scheme when $M>2$. More importantly, our results show that an optimal coding scheme for the symmetric Gaussian interference channels with feedback can be achieved by only using marginal posterior distributions under a better cooperation strategy between transmitters.                  
\end{abstract}   
\section {Introduction} \label{sec1}
The interference channels (IC) were first studied by Ahlswede~\cite{Ahlswede1974} in 1974, who established inner and outer bounds including the simultaneous decoding inner bound. Carleial~\cite{Carleial1974} introduced the idea of rate splitting and established an inner bound using successive cancellation decoding and time-sharing. His inner bound was improved through simultaneous decoding and coded time sharing by Han and Kobayashi~\cite{Han81}. For the two-user Gaussian interference channel as a special case, there have been some significant progresses toward finding better inner and outer bounds although the capacity of this channel has been open for nearly 40 years. \red{The approximation of the two-user Gaussian IC by a deterministic channel was first proposed by Bresler and Tse~\cite{Bresler2008}}. Furthermore, Etkin, Tse, and Wang~\cite{Etkin2008} proved that a very simple and explicit Han-Kobayashi type scheme can achieve the capacity for all values of the channel parameters within a single bit per second per hertz (bits/z/Hz).

Some other works have been done in the IC with feedback. Kramer developed a feedback strategy and derived an outer bound for the Gaussian channel. However, the gap between the outer bound and the inner bound becomes arbitrarily large with the increase of SNR (Signal-to-Noise Ratio) and INR (Interference-to-Noise Ratio)~\cite{Kramer2002a}. Jiang-Xin-Garg~\cite{Jiang2007e} found an achievable region in the discrete memoryless interference channel with feedback. However, their scheme employs three auxiliary random variables (requiring further optimization) and block Markov encoding (requiring a long block length). Suh and Tse~\cite{Suh2009e},~\cite{Suh2011a} characterized the capacity region within 2 bits/s/Hz and the symmetric capacity within 1 bit/s/Hz for the two-user Gaussian IC with feedback. They also indicated that feedback provides multiplicative gain at high SNR. However, their coding scheme does not work well when the SNR is close to the INR. It achieves even lower symmetric coding rate than the Kramer code when this condition happens. In addition, it has lower performance than the Kramer code when the $\alpha = \log INR/ \log SNR$ is not very large and the SNR is low (c.f. Figs. 2-3 of this paper, or Fig. 14 in~\cite{Suh2011a}). Recently, the Suh-Tse coding scheme \red{has been} extended to $M$-user Gaussian IC with feedback for $M \geq 3$~\cite{Tandon2013a},~\cite{Mohajer2013a} or the Gaussian IC with limited feedback~\cite{Ashraphijuo2014e}. The main ideas of these papers are to propose a method to manage the interference by turning the $M$-user Gaussian IC with feedback to an equivalent two-user one. Lattice codes, which are generally complicated in encoding and decoding, are used in these papers.  

In this paper, we propose a new coding scheme based on the Kramer code~\cite{Kramer2002a} and the time-varying posterior matching code~\cite{Truong2014e,TruongYamamoto2015e,TruongYamamoto17a,ShayevitzF}. Our code can attain better coding rate by using a devised transmission cooperation strategy and decoding only their intended messages based on the fact that the posterior distributions can be measured online at all transmitters and their corresponding receivers. The proposed coding scheme has the following strong points.
\begin{itemize}
\item {\bf Two-user case:} Our code can attain better symmetric coding rate than the Kramer code for all channel parameters since the Kramer code can be considered as a special case of our code and our code can be optimized more than the Kramer code for given channel parameters. Although the Kramer code cannot
achieve the generalized degrees-of-freedom (GDoF) of this IC~\cite{Suh2009e}, our code can attain the same GDoF as the Suh-Tse code~\cite{Suh2009e}. Furthermore, since our code can achieve better performance than the Suh-Tse code~\cite{Suh2009e} when $\alpha = \log INR/\log SNR$ is not very large (see Figs.~\ref{fig:HighSNRNew} and~\ref{fig:LowSNRnew} of this paper), our code overcomes all the weak-points of the Suh-Tse coding scheme and narrows the capacity gap to the Suh-Tse outer bound. 
\item {\bf $M$-user case for $M\geq 3$:} Our code can achieve the GDoF of the $M$-user symmetric Gaussian IC with feedback. Some numerical results show that our coding scheme can attain better performance than the Mohajer-Tandon-Poor lattice coding scheme~\cite{Mohajer2013a}, which achieves the GDoF for very weak and strong interferences. The good performance of our code comes from the use of marginal posterior distributions under a devised cooperation strategy between transmitters. For the special case such that the SNR is equal to the INR, our code includes the Kramer code as a special case. But we note that the Kramer code cannot  be constructed if the SNR is not equal to the INR for $M>2$. 
\end{itemize}  

\Red{In Section~\ref{sec2}, we describe the notation used in this paper and the channel model of the Gaussian IC. We propose our coding scheme for the Gaussian IC and evaluate the decoding error probability of the proposed code in Section~\ref{sec3}.  In Section~\ref{sec5}, we evaluate the normalized covariance matrix of channel inputs generated by our coding scheme. Then, we derive the symmetric coding rate of our code and also treat several special cases in Section~\ref{sec7}.  In Section~\ref{sec:num}, we show that our code can attain the GDoF of the Gaussian IC. Finally we show by numerical evaluations that our code can attain better symmetric coding rate than the Kramer code, the Suh-Tse code, and Mahajer-Tandon-Poor code.  }              
\section {Channel Model and Preliminaries} \label{sec2}
\subsection{Mathematical notations}\label{subsec21}
\red{Random variables and their realizations are denoted by upper-case letters and their corresponding lower-case letters, respectively.} A real-valued random variable $X$ is associated with a distribution $\mathbb{P}_X(\cdot)$ defined on the usual Borel $\sigma$-algebra over $\mathbb{R}$, and we write $X \sim \mathbb{P}_{X}$. The cumulative distribution function (c.d.f.) of $X$ is given by $F_X(x)=\mathbb{P}_X((-\infty,x])$, and their inverse c.d.f is defined to be $F_X^{-1}(t):= \mbox{inf}\{x:F_X(x) > t\}$. The uniform probability distribution over $(0,1)$ is denoted through $\mathcal{U}$. Then, it is known that the following lemma holds.

\begin{lemma}[{{\cite[Lemma 1]{ShayevitzF}}}] \label{ic:lem1} Let $X$ be a continuous random variable with $X\sim \mathbb{P}_X$ and $\Theta$ be a uniform distribution random variable which is statistically independent of $X$, i.e. $\Theta \sim \mathcal{U}$. Then $F_X^{-1}(\Theta) \sim \mathbb{P}_X$ and $F_X(X) \sim \mathcal{U}$.
\end{lemma}  

We also use the following notations: ${\bf Y}^{(n,m)}: = (Y_1^{(m)}, Y_2^{(m)},\ldots,Y_n^{(m)}), \log x:=\log_2(x)$, and $\exp_2(x):=2^x$. Landau's symbols $O(\cdot)$ and $o(\cdot)$ are defined as follows.
\begin{align*}
f(n)=O(g(n))
\end{align*} if and only \red{if there exists real positive constants $N$ and $C$} such that
\begin{align*}
|f(n)| \leq C|g(n)| \quad \mbox{for all} \quad n >N.
\end{align*} 
Intuitively, this means that $f$ does not grow faster than $g$. 
\begin{align*}
f(n)=o(g(n))
\end{align*} if and only if there exists a real number $N$ for any $C>0$ such that $|f(n)| < C |g(n)|$ for all $n>N$. If $g(n) \neq 0$, this is equivalent to $\lim_{n\rightarrow \infty} f(n)/g(n)=0$.
 
A Hadamard matrix~\cite{Wicker} of order $M$ is an $(M\times M)$ matrix of $+1$s and $-1$s such that $\bH\bH^T=M \bI_M$. In fact, it is not yet known for which values of $M$ an $\bH$ exists. However, we know that if a Hadamard matrix of order $M$ exists, then 
\begin{figure}[!tbp] 
  \centering
  \begin{minipage}[t]{0.40\textwidth}
    \includegraphics[width=\textwidth]{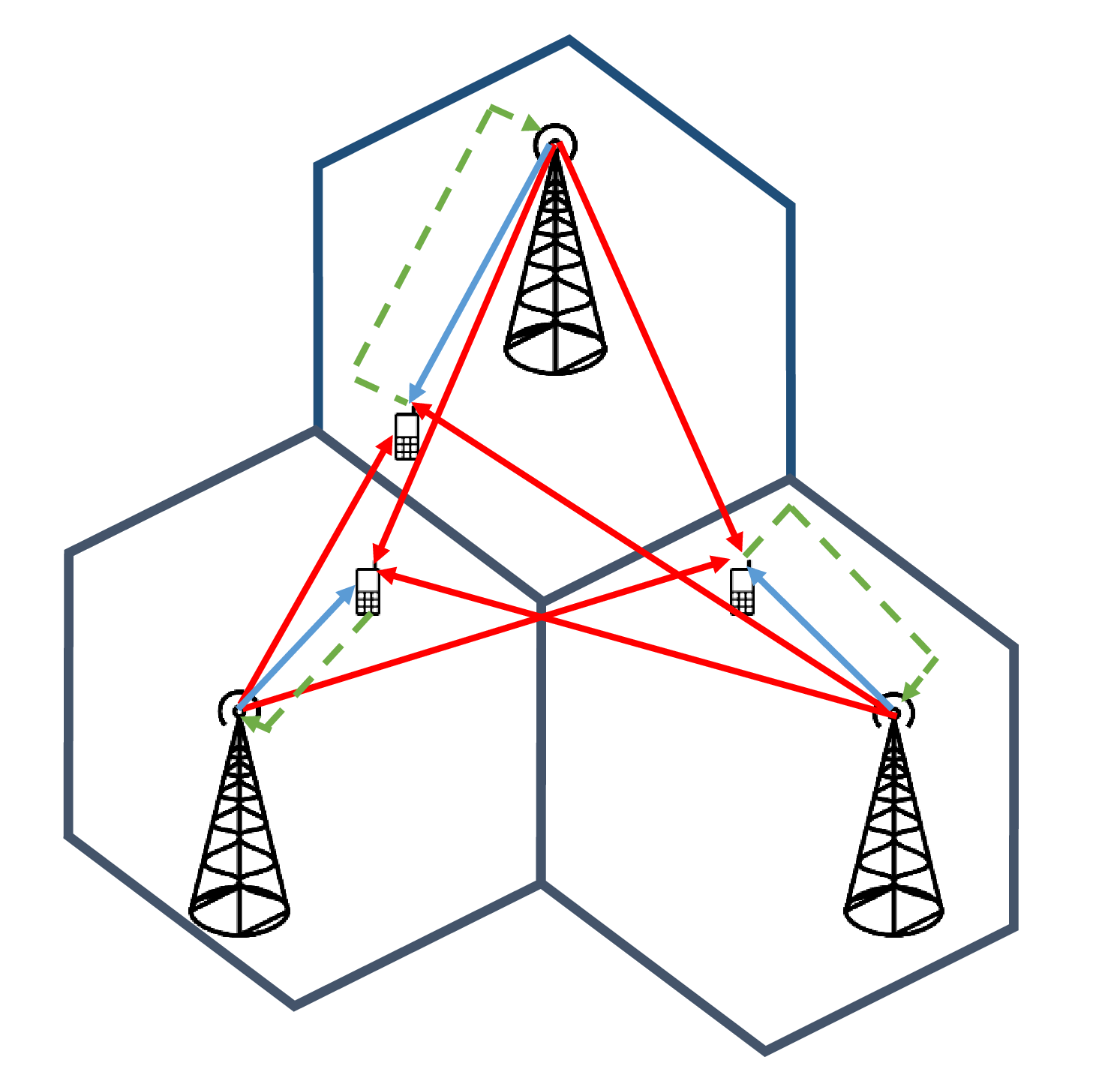} 
		\label{sub1:tradeoff}
  \end{minipage}
	\hfill
   \begin{minipage}[t]{0.40\textwidth}
    \includegraphics[width=\textwidth]{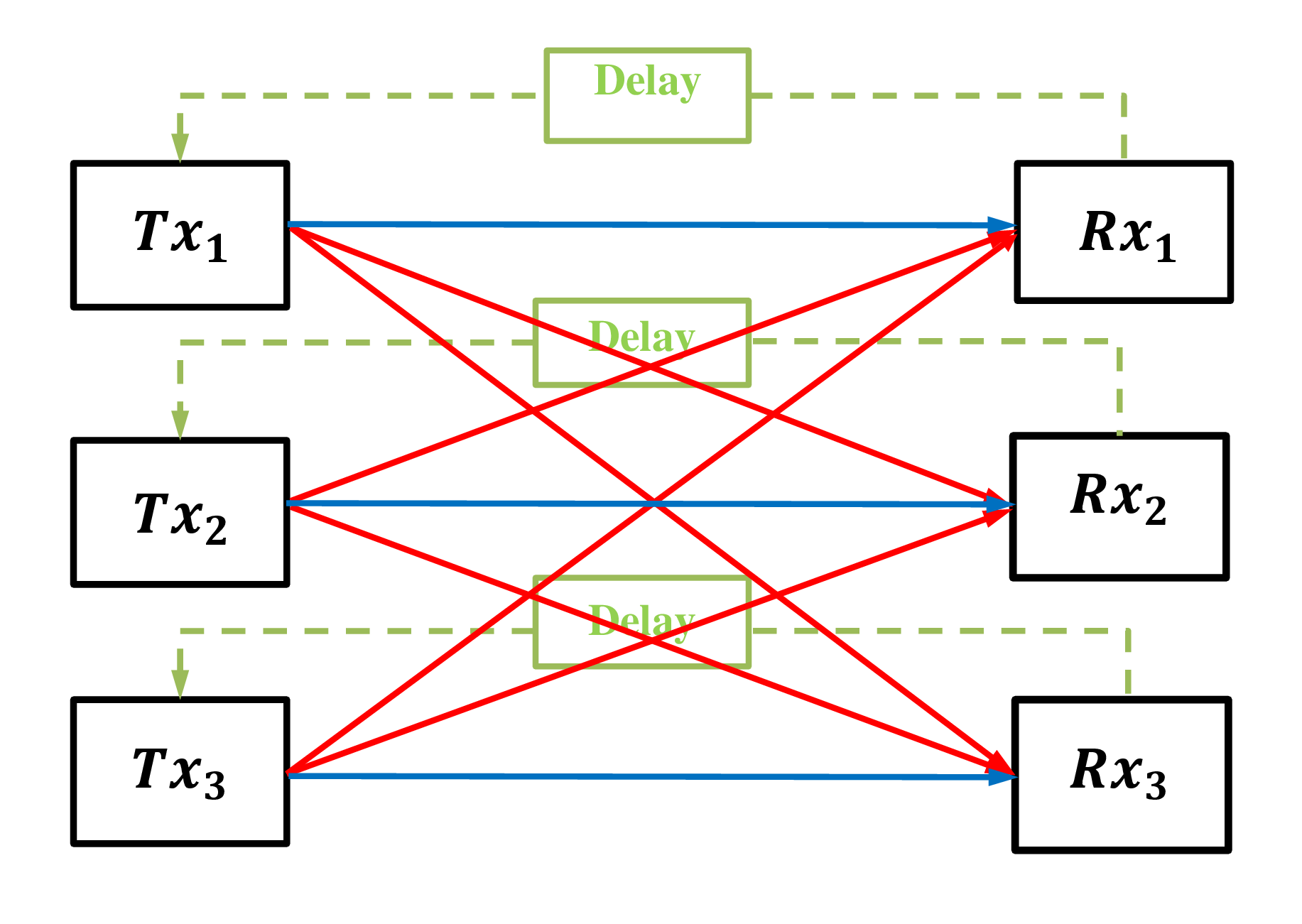}
  \end{minipage}
	\caption{Cellular network with base stations and three clients in (a), simplified and modeled as the network in (b).}
	\label{fig:tradeoff}
\end{figure}
$M$ is $1, 2, 4$, or a multiple of $4$. Moreover, for $M=2^m$ where $m$ a positive integer,  we can construct $\bH$ by using the Sylvester method [19]. Besides,  the Paley construction~\cite{Wicker}, which uses quadratic residues, can be used to construct Hadamard matrices of order $M$ if $M=p+1$ for a prime $p$ and $M$ is a multiple of $4$. 

\Red{Let ${\boldsymbol \alpha}_n$ be the $[(n-1\mod M)+1]$-th column of the Hadamard matrix $\bH$. In this paper, we use column-permutation matrices of the Hadamard matrix $\bH$, say $\bH_n, n \in \bbZ^+$, which are defined as follows:  
\begin{align}
\label{defHn}
\bH_n:=\big[\begin{array}{cccc} {\boldsymbol \alpha}_n &{\boldsymbol \alpha}_{n+1}& \cdots & {\boldsymbol \alpha}_{n+M-1} \end{array}\big].
\end{align}}

\subsection{Gaussian Interference Channel with Feedback}\label{subsec22}
Consider a network with $M$ pairs of transmitters/receivers shown in Fig. 1(b). Each transmitter ${\bf Tx}_{m}$ has a message $\Theta_m \sim \mathcal{U}(0,1)$ that it wishes to send to its respective receiver ${\bf Rx}_{m}$. The signal transmitted by each transmitter is corrupted by the interfering signals sent by other transmitters, and received at the receiver. This can be mathematically modeled as
\begin{align}
\label{ic:eq1}
Y^{(m)}_n=X^{(m)}_n + a\sum_{k=1, k\neq m}^M X_n^{(k)} + Z_n^{(m)}
\end{align}
where $X^{(m)}_n \in \mathbb{R}$ is the transmitted symbol by sender $m$ at time $n$; $Y^{(m)}_n \in \mathbb{R}$ is the received signal by receiver $m$ at the time $n$. We can assume without loss of generality $Z_n^{(m)}\sim \mathcal{N}(0,1)$ and $a\geq 0$.  We also assume that the output symbols are casually fed back to the corresponding senders and that the transmitted symbol $X^{(m)}_n$ at time $n$ can depend on both the message $\Theta_m$ and the previous channel output sequences ${\bf Y}^{(n-1,m)}:=\left(Y^{(m)}_1,Y^{(m)}_2,\ldots, Y^{(m)}_{n-1}\right),\hspace{2mm} \forall m \in \{1,2,\ldots,M\}$. 

A \emph{transmission scheme} for the $M$-user Gaussian interference channel with feedback is sequences of measurable functions $\{g^{(m)}_n: (0,1) \times \mathbb{R}^{n-1} \rightarrow \mathbb{R}\}_{n=1}^{\infty}, m\in\{1,2,\ldots,M\}$ so that the input to the channel generated by the transmitter is given by
\begin{align}
\label{ic:eq2}
X^{(m)}_n = g^{(m)}_n(\Theta_m,{\bf Y}^{(n-1,m)}).
\end{align}

A \emph{decoding rule} for the $M$-user Gaussian interference channel with feedback are sequences of measurable mappings $\{\Delta_n^{(m)}: \mathbb{R}^n \rightarrow \mathcal{E}\}_{n=1}^{\infty}, m\in \{1,2,\ldots,M\}$ where $\mathcal{E}$ is the set of all open intervals in $(0,1)$ and $\Delta_n^{(m)}({\bf y}^{(n,m)})$ is referred to as the decoded interval at receiver $m$. The error probabilities at time $n$ associated with a transmission scheme and a decoding rule, is defined as
\begin{align}
\label{ic:eq3}
p_n^{(m)}(e):=\mathbb{P}(\Theta_m \notin \Delta_n^{(m)}({\bf Y}^{(n,m)})), \quad \forall m=1,2,\ldots,M,
\end{align}
and the corresponding coding rate vector $(R_n^{(1)}, R_n^{(2)},\ldots,R_n^{(M)})$ at time $n$ is defined by
\begin{align}
\label{ic:eq4}
 R_n^{(m)}:=-\frac{1}{n}\log\left|\Delta_n^{(m)}\left({\bf Y}^{(n,m)}\right)\right|,
\end{align}
where $|\cdot|$ represents the length of an interval.

We say that a transmission scheme together with a decoding rule achieves a rate vector $(R_1, R_2,\ldots,R_M)$ over a Gaussian interference channel if for all $m\in \{1,2,\ldots,M\}$ we have
\begin{align}
\label{ic:eq5}
\lim_{n\rightarrow \infty}\bbP\left(R^{(m)}_n<R_m\right)&=0,\\
\label{ic:eq6}
\lim_{n \rightarrow \infty}p_n^{(m)}(e)&=0.
\end{align}
The rate vector is achieved within input power constraints $P^{(1)}, P^{(2)},\ldots,P^{(M)}$ if the following is satisfied:
\begin{align}
\label{ic:eq7}
\limsup_{n\rightarrow \infty}\frac{1}{n}\sum_{k=1}^n \bbE[(X^{(m)}_k)^2] \leq P^{(m)}, \quad \forall m=1,2,\ldots,M.
\end{align}
We denote the set of all achievable rate tuples $(R_1,R_2,\ldots,R_M)$ by $\calR$. 

For the symmetric case~\cite{Mohajer2013a}, i.e. $P^{(1)}=P^{(2)}=\cdots=P^{(M)}=P$ for some $P>0$, let
\begin{align}
SNR&:=P,\\
INR&:=a^2P,\\
\alpha&:=\frac{\log INR}{\log SNR},
\end{align}
and define the per-user generalized degrees of freedom for $R_m=R_m(SNR,\alpha)$ as
\begin{align}
\label{deg}
d(\alpha)=\frac{1}{M}\limsup_{SNR \to \infty} \frac{\max_{(R_1,R_2,\ldots,R_M) \in \calR}\sum_{m=1}^M R_m(SNR,\alpha)}{(1/2)\log (SNR)}.
\end{align}
If $R_1=R_2=\cdots=R_M=R_{\rm{sym}}$ we call $R_{\rm{sym}}$ a symmetric rate, and the symmetric capacity is defined by 
\begin{align}
\label{ic:eq8}
C_{\mathrm{sym}}:= \sup\{R_{\rm{sym}}: (R_{\rm{sym}},R_{\rm{sym}},\ldots,R_{\rm{sym}}) \in \calR\}.
\end{align}
The per-user generalized degrees of freedom in~\eqref{deg} can be written as
\begin{align}
d(\alpha)=\limsup_{SNR \to \infty}\frac{C_{\rm{sym}}}{(1/2)\log (SNR)}.
\end{align}
An \emph{optimal fixed rate} decoding rule for the $M$-user Gaussian interference channel with feedback for rate vector $(R_1,R_2,\ldots,R_M)$ is the one that decodes a vector of fixed length intervals $\{(J_1,J_2,\ldots,J_M): |J_m|=2^{-nR_m} \hspace{1mm}\mbox{for}\hspace{1mm} m\in\{1,2,\ldots,M\}\}$, which maximizes posteriori probabilities, i.e.,
\begin{align}
\label{ic:eq9}
\triangle^{(m)}_n({\bf y}^{(n,m)})=\underset{{J_m \in \mathcal{E}: |J_m|=2^{-nR_m}}}{\mbox{argmax}} \bbP_{\Theta_m|Y^n}(J_m|{\bf y}^{(n,m)}).
\end{align}
It is easy to see that the optimal fixed rate decoding rule for the Gaussian interference channel with feedback is the traditional MAP, MMSE decoding rule.

An \emph{optimal variable rate} decoding rule with target error probabilities $p^{(m)}_e(n)=\delta^{(m)}_n$ is the one that decodes a vector of minimal-length intervals $(J_1,J_2,\ldots,J_M)$ such that accumulated marginal posteriori probabilities exceeds corresponding targets, i.e.,
\begin{align}
\label{ic:eq10}
\triangle^{(m)}_n(y^{(n,m)})=\underset{{J_m\in \mathcal{E}: \mathbb{P}_{\Theta_m|Y^n}(J_m|y^{(n,m)})\geq 1- \delta^{(m)}_n}}{\mbox{min}}|J_m|.
\end{align}
Both decoding rules use the marginal posterior distribution of the message point $\mathbb{P}_{\Theta_m|{\bf Y}^n}$ which can be calculated online at the transmitters and the receivers.  Refer~\cite{Truong2014e,TruongYamamoto2015e, TruongYamamoto17a}, and~\cite{ShayevitzF} for more details. \\

\begin{lemma}\label{ic:lem2}
 The achievability in the definition~\eqref{ic:eq5},~\eqref{ic:eq6}, and~\eqref{ic:eq7} implies the achievability in the standard framework.
\end{lemma}
\begin{IEEEproof}
\red{See the detailed proof in papers~\cite{ShayevitzF},~\cite{Truong2014e}. The idea is as follows. Assume that we use an encoding scheme $\{g_n^{(m)}\}_{m=1}^M$ and a decoding rule $\{\Delta_n^{(m)}\}_{m=1}^M$ to achieve a rate tuple $\{R^*_m\}_{m=1}^M$ under the assumption that all the transmitted messages $\Theta_m$ are mutually independent and uniformly distributed on $(0,1)$ as in Section~\ref{subsec22}. Then, we can show the existence of $M$ sequences $\Gamma_m=\{\theta_{i,n}^{(m)} \in (0,1)\}_{i=1}^{\lfloor 2^{nR_m}\rfloor}$ of message point sets for $m=1,2,\ldots,M$, where any two message points in $\Gamma_m$ are separated from each other by at least $2^{-nR^*_m}$.   In addition, a uniform achievability over $\{\Gamma_m\}_{m=1}^M$ is guaranteed, i.e., $\lim_{n\rightarrow \infty} \max_{\theta_n^{(m)} \in \Gamma_m} \mathbb{P}(\theta_n^{(m)} \notin \Delta_n^{(m)}({\bf Y}^{(n,m)})|\Theta_m=\theta_n^{(m)})=0$ for each $m\in \{1,2,\ldots,M\}$. By mapping message points $\{1, 2,\ldots, \lfloor 2^{nR_m^*}\rfloor\}$ defined in the traditional way\footnote{In the traditional settings, we usually assume that message set at user $m$ is $\{1,2,\ldots, 2^{nR_m}\}$, where $R_m$ is an achievable rate of the code~\cite{Cov06}.} to message points in $\Gamma_m$, each coding scheme in Section~\ref{subsec22} is reduced to an equivalent coding scheme in the traditional setting as in~\cite{Cov06}.  The error probabilities of the associated scheme decay as $\sqrt{p_n^{(m)}(e)}$~\cite[Proof of Lemma II.3]{ShayevitzF}. }
\end{IEEEproof}
\section{A General Time-varying Coding Scheme for The Symmetric Gaussian Interference  Channel with Feedback}\label{sec3}
In this section, we propose a \emph{time-varying encoding/decoding} scheme for the \emph{symmetric} Gaussian interference channel with feedback. For this symmetric case, we assume that $P^{(1)}=P^{(2)}=\cdots=P^{(M)}=P$ for some $P>0$. The time-varying encoding scheme is as follows:
\subsection{Encoding}\label{subsec31}
\begin{itemize}
\item Step 1:
Transmitter $m$ sends $X^{(m)}_1=F_{X}^{-1}(\Theta_m),\hspace{1mm} m\in \{1, 2,\ldots,M\}$, where $X \sim \mathcal{N}(0,P_1)$ for some $P_1 >0$. 
\item Step $n+1,\hspace{2mm} n \geq 1$:\\
\begin{itemize}
\item All transmitters estimate
\begin{align}
\label{ic:eq12}
&P_{n+1}=\frac{P_n}{\beta_n^2}\bigg( [1-b_n(1-a)]^2 \nn\\
&\qquad + ab_n\lambda_n[2(1-a)b_n+Mab_n-2] +\frac{b_n^2}{P_n}\bigg),
\end{align} 
and each transmitter $m$ sends
\begin{align}
\label{ic:eq14}
X_{n+1}^{(m)} \alpha_{n+1}^{(m)}:=\frac{1}{\beta_n}(X_n^{(m)}- b_n \alpha_n^{(m)} Y_n^{(m)})\alpha_{n+1}^{(m)},
\end{align} where ${\boldsymbol \alpha}_{n+1}:=[\begin{array}{cccc}\alpha_{n+1}^{(1)}&\alpha_{n+1}^{(2)}&\ldots &\alpha_{n+1}^{(M)}\end{array}]^T$, $\alpha^{(m)}_{n+1}\in\{-1, 1\}$, is the $\left((n \mod  M)+1\right)$-th column of an $M\times M$ Hadamard matrix $\bH$ which is defined in Section~\ref{subsec21}, and $b_n$ and $\beta_n$ are some real number sequences such that $0<\limsup_{n\to \infty}\beta_n<1$ and they are
determined according to network situations.
\item Receiver $m$ receives
\begin{align}
\label{ic:eq15}
Y_{n+1}^{(m)} = X_{n+1}^{(m)}\alpha_{n+1}^{(m)} + a\sum_{k=1, k\neq m}^M \alpha_{n+1}^{(k)} X_{n+1}^{(k)} + Z_{n+1}^{(k)},
\end{align}
and each receiver feedbacks the received signal to the corresponding transmitter.
\end{itemize}
Here, $\{P_1,\beta_n, b_n\}$ must be chosen to satisfy the following power constraint:
\begin{align}
\label{ic:eq16}
\limsup_{N\rightarrow \infty} \frac{1}{N} \sum_{n=1}^N \bbE[(X_n^{(m)})^2] \leq P, \quad \forall m\in\{1,2,\ldots,M\}.
\end{align}
\end{itemize}
\red{\begin{remark} \label{addedrmk} The outperform of time-varying codes over the Kramer codes~\cite{Kramer2002a} can be explained by our better choice of parameter triplet $(P_1,\beta_n,b_n)$ for each network situation. We note that each triplet of parameter $(P_1,\beta_n,b_n)$ represents a cooperation strategy. With better cooperation among all transmitters, we can achieve larger achievable rate region. Besides, the use of (real) Hadamard matrix as coefficient matrix allows us to analyze and optimize the performance easier than the use of (complex) DFT matrix for the same purpose as in~\cite{Kramer2002a}. (Refer to Corollary~\ref{lemconff} in Section~\ref{2user} for more details.)
\end{remark}}
\subsection{Decoding}\label{subsec32}
 \begin{itemize}
\item At each time slot $n$, receiver $m \in \{1,2,\ldots,M\}$ selects a fixed interval $J_1^{(m)}=(s_m,t_m)$ as the decoded interval with respect to $X_n^{(m)}$. \Red{Here,
\begin{align}
\label{deftm}
t_m&:=o(2^{n(\log(\beta+\eps)^{-1}-R)}),\\
\label{defsm}
s_m&:=-t_m,
\end{align}
where $\beta:=\limsup_{n\to \infty} \beta_n$, and $R$ is any positive number such that $R<\log (\beta+\eps)^{-1}$ for some $\eps>0$.
}
\item Then, set $J_n^{(m)}=\big(T^{(m)}_{n-1}(s_m),T^{(m)}_{n-1}(t_m)\big)$ as the decoded interval with respect to $X_1^{(m)}$, where  
 \begin{align}
\label{ic:eq17}
 T^{(m)}_n(s):= w^{(m)}_1\circ w^{(m)}_2\circ \cdots \circ w^{(m)}_n (s), \hspace{2mm} \forall s\in \mathbb{R}
 \end{align}
and 
\begin{align}
\label{ic:eq18}
w^{(m)}_n(s):=\beta_n s + b_n \alpha_n^{(m)} Y_n^{(m)},
\end{align} \Red{where $\circ$ is the composition operation defined as $(f \circ g)(x):=f(g(x))$.}
\item The receiver $m$ sets the decoded interval for the message $\Theta_m$ as
\begin{align}
\label{ic:eq19}
\Delta_n^{(m)}\left({\bf Y}^{(n,m)}\right)=F_X(J_n^{(m)}),
\end{align} where $X \sim \calN(0,P_1)$.
\end{itemize}
We call this coding strategy the \emph{Gaussian interference time-varying feedback coding strategy}, which is an \emph{optimal variable rate} decoding rule with doubly exponential decay of targeted error probabilities  (see the proof of the Theorem~\ref{ic:thm1} in this paper).
\subsection{Analysis of decoding error probability}\label{sec4}
In this subsection, we evaluate the performance of the \emph{Gaussian interference time-varying feedback coding strategy} defined in the above subsections.
\begin{theorem}\label{ic:thm1} Under the condition that $0< \beta:=\limsup_{n\rightarrow \infty} \beta_n < 1$ and $W:=\sup_{n,m} \bbE[(X_n^{(m)})^2]< \infty$, the \emph{time-varying coding scheme}  for the symmetric Gaussian interference channel with feedback achieves the following symmetric rate:
\begin{align}
\label{ic:eq20}
R_{\rm{sym}} = \log \frac{1}{\beta}  \quad  \text{(bits/channel use)}.
\end{align}
\end{theorem}
\begin{IEEEproof} 
Define
\begin{align}
f_X(x)&:=\frac{1}{\sqrt{2\pi P_1}} \exp\left(-\frac{x^2}{2P_1}\right),\\
\label{Ksup:ic}
K &:=\sup_{x \in \mathbb{R}}\{f_X(x)\}=\frac{1}{\sqrt{2\pi P_1}}.
\end{align}
Let $R_n^{(m)}$ be the instant rate to transmit the intended messages $\Theta_m$ to the receiver $m$. For any fixed symmetric rate $R$, we have
\begin{align}
\bbP\left(R_n^{(m)}< R \right) &\stackrel{(a)}{=}\bbP \left(-\frac{1}{n} \log\left|\Delta_n^{(m)}({\bf Y}^{(n,m)})\right|  < R\right)  \nonumber\\
&=\bbP \left(|\Delta_n^{(m)}({\bf Y}^{(n,m)})| > 2^{-nR}\right)  \nonumber \\
&\stackrel{(b)}{=}\bbP\bigg(\int_{J_n^{(m)}} f_X(x)dx >2^{-nR}\bigg) \nonumber \\
\label{ic:eq23}
&\stackrel{(c)}{\leq} \bbP \left( |J_n^{(m)}|> 2^{-n R}/ K\right),
\end{align}
where~(a) follows from~\eqref{ic:eq4},~(b) follows from~\eqref{ic:eq19}, and~(c) follows from~\eqref{Ksup:ic}. In addition, it holds from~\eqref{ic:eq18} that for any $t, s \in \mathbb{R}$
\begin{align}
\label{aquafina1}
|w_n^{(m)}(t)- w_n^{(m)}(s)| = \beta_n |t-s|
\end{align} for all $ m\in \{1,2,\ldots,M\}$ and $n=1,2,\ldots$.

Now, we will show that rate $R$ is achievable if $R<R_{\rm{sym}}=\log\beta^{-1}$. Since $R < R_{\rm{sym}}$, we can find an $\epsilon >0$ such that $R < \log (\beta +\epsilon)^{-1}$. Therefore, for $n$ sufficiently large, we have
\begin{align}
&\bbP\left(R_n^{(m)}< R \right)\nn\\
&\qquad \stackrel{(a)}{\leq} K 2^{nR} \bbE\bigg[\big|w_1^{(m)}\circ w_2^{(m)} \cdots \circ w_{n-1}^{(m)}(t_m)\nn\\
&\qquad  \qquad -w_1^{(m)}\circ w_2^{(m)} \cdots \circ w_{n-1}^{(m)}(s_m)\big|\bigg] \nn \\
&\qquad \stackrel{(b)}{=} K 2^{nR} \bbE\bigg[\bbE\bigg(|w_1^{(m)}\circ w_2^{(m)} \cdots \circ w_{n-1}^{(m)}(t_m) \nn\\
&\qquad \qquad -w_1^{(m)}\circ w_2^{(m)} \cdots \circ w_{n-1}^{(m)}(s_m)|\bigg)\Big|\bY_2^{(n-1,m)}\bigg] \nn \\
&\qquad \stackrel{(c)}{=} K 2^{nR} \beta_1 \bbE\bigg[\bbE\bigg(|w_2^{(m)}\circ w_3^{(m)} \cdots \circ w_{n-1}^{(m)}(t_m) \nn\\
&\qquad \qquad -w_2^{(m)}\circ w_3^{(m)} \cdots \circ w_{n-1}^{(m)}(s_m)|\bigg)\Big| \bY_2^{(n-1,m)}\bigg] \nn \\
&\qquad \stackrel{(d)}{= } K 2^{nR} \beta_1 \bbE\bigg[|w_2^{(m)} \circ w_3^{(m)} \cdots \circ w_{n-1}^{(m)}(t_m) \nn\\
&\qquad \qquad -w_2^{(m)} \circ  w_3^{(m)} \cdots \circ w_{n-1}^{(m)}(s_m)|\bigg] \nn \\
&\qquad \qquad \vdots  \nonumber \\
&\qquad \stackrel{(e)}{=} K 2^{nR}\bigg(\prod_{i=1}^{n-1} \beta_i\bigg) \bbE\bigg[|w_n^{(m)}(t_m)
-w_n^{(m)}(s_m)|\bigg] \nonumber \\
\label{ic:eq31}
&\qquad \stackrel{(f)}{=} K 2^{nR}\bigg(\prod_{i=1}^n \beta_i\bigg)|J_1^{(m)}| \nn \\
&\qquad = K 2^{nR}\big(2^{\frac{1}{n}\sum_{i=1}^n \log \beta_i}\big)^n |J_1^{(m)}| \nn \\
&\qquad \stackrel{(g)}{\leq } K 2^{nR} 2^{n\log (\beta+\eps)} |J_1^{(m)}|,
\end{align}
where~(a) follows from the Markov's inequality and~\eqref{ic:eq23},~(b) and~(d) follow from the law of iterated expectations,~(c) follows from~\eqref{aquafina1} for each fixed $\bY_2^{(n,m)}=\by_2^{(n,m)}$,~(e) follows from the recursive application of~(b)-(d),~(f) follows from~\eqref{aquafina1}, and~(g) follows from $0< \beta_n < \beta+\eps $ and Ces\`{a}ro means for $n$ sufficiently large.

\Red{From~\eqref{deftm},~\eqref{defsm}, and~\eqref{ic:eq31}, it is easy to see that $\mathbb{P}(R_n^{(m)} < R) \rightarrow 0$ since
\begin{align}
\label{j1mchoice}
|J_1^{(m)}| = 2t_m=o\left(2^{n(\log(\beta+\epsilon)^{-1} - R)}\right).
\end{align}}

Now, from our encoding scheme, it is easy to see that
\begin{align}
\label{ic:easyfact}
\bbE[X_n^{(m)}]=0, 
\end{align} for all $m\in \{1,2,\ldots,M\}$ and $n=1,2,\ldots$ \Red{On the other hand, observe that
\begin{align}
X_n^{(m)}&\stackrel{(a)}{=}\beta_n X_{n+1}^{(m)} + b_n \alpha_n^{(m)} Y_n^{(m)} \nonumber\\
\label{sup2}
&\stackrel{(b)}{=}w_n^{(m)}( X_{n+1}^{(m)})
\end{align} for all $n\geq 1$ and $m \in \{1,2,\cdots,M\}$, where~(a) and~(b) follow from~\eqref{ic:eq14} and~\eqref{ic:eq18}, respectively. Therefore, we have from~\eqref{sup2} that
\begin{align}
X_1^{(m)}&=w_1^{(m)}(X_2^{(m)})\nonumber\\
&=(w_1^{(m)}\circ w_2^{(m)})(X_3^{(m)})\nonumber\\
&\quad \vdots \nonumber\\
&=(w_1^{(m)}\circ w_2^{(m)}\circ w_3^{(m)} \cdots \circ w_{n-1}^{(m)})(X_n^{(m)}) \nonumber\\
\label{sup4}
&= T_{n-1}^{(m)}(X_n^{(m)}),
\end{align} where the last equality in~\eqref{sup4} follows from~\eqref{ic:eq17}.
 }

Let $W_n^{(m)}=\bbE[(X_n^{(m)})^2]$, and let $Q(x)$ as the well-known tail function of the standard normal distribution $\mathcal{N}(0,1)$. Then, from the Chernoff bound of this function, we obtain
\begin{align}
p_n^{(m)}(e) &=\bbP\left(\Theta_m \notin \Delta_n^{(m)}\left({\bf Y}^{(n,m)}\right)\right) \nonumber \\
&=\bbP\left(\Theta_m \notin F_X(J_n^{(m)})\right) \nonumber \\
&=\bbP\left(X_1^{(m)} \notin J_n^{(m)}\right) \nonumber \\
&\Red{\stackrel{(a)}{=} \bbP\left(T_{n-1}^{(m)}(X_n^{(m)}) \notin \bigg(T_{n-1}^{(m)}(s_m),T_{n-1}^{(m)}(t_m)\bigg)\right)} \nonumber \\
&= \bbP\left(X_n^{(m)} \notin J_1^{(m)}\right) \nonumber \\
&\stackrel{(b)}{=} 2Q\left(\frac{|J_1^{(m)}|}{2\sqrt{W_n^{(m)}}}\right) \nonumber \\
\label{mac:eq34}
&\stackrel{(c)}{\leq } \exp\left(-\frac{|J_1^{(m)}|^2}{8W_n^{(m)}}\right),
\end{align}
where~(a) follows from~\eqref{sup4} and the definition of $J_n^{(m)}$,~(b) follows from~\eqref{ic:easyfact} and the fact that $J_1^{(m)}$ is symmetric and $s_m= -t_m$, and~(c) follows from the Chernoff bound for the Q-function $0 < Q(x) \leq (1/2) \exp(-x^2/2), \forall x >0$. 

From $R <\log(\beta+\epsilon)^{-1} < R_{\rm{sym}}$, we have that for each $m\in \{1,2,\ldots,M\}$, $|J_1^{(m)}| \to \infty$ as $n\to \infty$ . Furthermore, from the assumption that $W_n^{(m)} $ is upper bounded by some $W$, we have
\begin{align}
\frac{|J^{(m)}_1|^2 }{8W^{(m)}_n} \geq \frac{|J^{(m)}_1|^2 }{8W} \rightarrow \infty
\end{align} as $n\to \infty$.
Therefore, if $R <R_{\mathrm{sym}}$, the error probabilities tend to zero as 
$
-\log p_n^{(m)} (e) = o(2^{2n(\log(\beta+\epsilon)^{-1} -R)})$ from~\eqref{j1mchoice} and~\eqref{mac:eq34}. Furthermore, since $(P_1, b_n, \beta_n)$ is chosen to satisfy~\eqref{ic:eq16}, the input power constraints are also satisfied.
\end{IEEEproof} 
\begin{remark}
Since we can estimate $R_{\mathrm{sym}}$ and know our desired rate $R$ in advance, it is possible to choose $\epsilon$ appropriately. This means that the decoding algorithm can be implemented practically.  However, there is a tradeoff between the transmission rate $R$ (the possible values of $\epsilon$) and the code length $n$. If we transmit at a rate $R$ very close to $R_{\mathrm{sym}}$, $\epsilon$ must be very small. As a result, the required $N_{\epsilon}$ may become very large. Furthermore, the fact that $\log (\beta +\epsilon)^{-1}$ is very close to $R$ also implies that the error probabilities decay slowly to zero. Therefore, the code length $n$ must be very large if we transmit nearly at $R_{\mathrm{sym}}$. On the contrary, quite large $\epsilon$ makes the required $N_{\epsilon}$ smaller and the decay of error probabilities faster. 
\end{remark}
\begin{remark} \label{rmk-re}
In the case of finite $n$,  $\mathbb{P}(R_n^{(m)}< R)$ is not zero even if $J_1^{(m)}$ satisfies~\eqref{j1mchoice}. But this does not worsen the error probability $p^{(m)}_n(e)$ if  retransmission is allowed. Note that since each encoder $m$ obtains ${\bf y}^{(n,m)}$ via the feedback channel, both encoder $m$ and decoder $m$ can know the value of $R_n^{(m)}$ for ${\bf y}^{(n,m)}$. Hence, they can know whether event $\{R_n^{(m)}< R\}$ occurred or not for received ${\bf y}^{(n,m)}$. If  event $\{R_n^{(m)}< R\}$ occurs, they discard this transmission and resend the same message $\Theta_m$.
This retransmission decreases the coding rate  of message $\Theta_m$  from $R_n^{(m)}$ to $R_n^{(m)} (1 - \mathbb{P}(R_n^{(m)}< R))$. But, this degradation of coding rate is negligible if $\mathbb{P}(R_n^{(m)}< R)$ is sufficiently small. 
\end{remark}
\begin{remark}
If we cannot use the retransmission described in Remark \ref{rmk-re}, event $\{R_n^{(m)}< R\}$ makes a decoding error. In this case, we need to minimize the total decoding error probability given by 
$p_n^{(m)}(e)+ \mathbb{P}(R_n^{(m)}< R)$, and hence we cannot attain double exponential order. By setting $|J_1^{(m)}|^2(\log {\rm e})/8W= n(\log(r^{(m)}+\epsilon)^{-1}-R)$ in \eqref{mac:eq34},  the error exponent of the total error probability is given by
\begin{align}
\lim_{\epsilon\rightarrow0}\lim_{n\rightarrow\infty}&\left[-\frac{1}{n} \log \left(p_n^{(m)}(e)+ \mathbb{P}(R_n^{(m)}< R)\right)\right] \nonumber\\
 &\geq\lim_{\epsilon\rightarrow0}\left(\log(r^{(m)}+\epsilon)^{-1}-R\right)\nonumber\\
 &= R_{\mathrm{sym}}-R.\label{eq-new32}
\end{align}
\end{remark}
\section{Normalized Covariance Matrix of channel inputs for the proposed scheme}\label{sec5}
Firstly, we show the following propositions.
\begin{proposition}\label{ic:pro1}
For $\bbE[(X_n^{(1)})^2]=\bbE[(X_n^{(2)})^2]=\cdots=\bbE[(X_n^{(M)})^2]:=P_n$, define a normalized covariance matrix by
\begin{align}
{\bf R}_n&:=\frac{1}{P_n}\left[ \begin{array}{cccc}\bbE[X_n^{(1)} X_n^{(1)}] &\cdots&\bbE[X_n^{(1)}X_n^{(M)}]\\ \bbE[X_n^{(2)}X_n^{(1)}] &\cdots&\bbE[X_n^{(2)}X_n^{(M)}]\\ \vdots&\ddots&\vdots\\ \bbE[X_n^{(M)}X_n^{(1)}] &\cdots& \bbE[X_n^{(M)}X_n^{(M)}]\end{array}\right]  \nonumber \\
\label{ic2018:eq42}
&=\left[ \begin{array}{cccc}\rho^{(1,1)}_n&\cdots&\rho^{(1,M)}_n\\\rho^{(2,1)}_n&\cdots&\rho^{(2,M)}_n\\ \vdots&\ddots&\vdots\\\rho^{(M,1)}_n &\cdots&\rho^{(M,M)}_n\end{array}\right],
\end{align}
where  
\begin{align}
\label{ic:eq41}
\rho^{(m,k)}_n&:=\frac{\bbE[X_n^{(m)}X_n^{(k)}]}{P_n},\\
\label{ic:eq42}
\rho^{(m,k)}_n&=\rho^{(k,m)}_n, \quad \rho^{(m,m)}_n=1
\end{align} for all $m=1,2,\ldots,M$ and $k=1,2,\ldots,M$, then the following statement holds:
 
If the covariance matrix ${\bf R}_n$ at time $n$ has all the columns of the $M \times M$ Hadamard matrix as its eigenvectors, it follows that $\bbE[(X_{n+1}^{(1)})^2]=\bbE[(X_{n+1}^{(2)})^2]=\cdots=\bbE[(X_{n+1}^{(M)})^2]:=P_{n+1}$ and the covariance matrix ${\bf R}_{n+1}$ at time $n+1$ also has all the columns of the $M \times M$ Hadamard matrix as its eigenvectors. In addition,~\eqref{ic:eq66} holds,
\begin{figure*}
\begin{equation}
\label{ic:eq66}
P_{n+1}\lambda_{n+1}^{(k)}=\begin{cases}\frac{P_n}{\beta_n^2}\left([1-b_n(1-a)]^2\lambda_n^{(k+1)}+\frac{b_n^2}{P_n}\right),&\qquad k=1,2,\ldots,M-1 \\ 
\frac{P_n}{\beta_n^2}\bigg([1-b_n(1-a)]^2\lambda^{(1)}_n +\frac{b_n^2}{P_n}+ ab_n\lambda_n^{(1)}[2(1-a)b_n + M a b_n -2]M \bigg),& \qquad k=M\end{cases},
\end{equation}\hrulefill
\end{figure*}
where $\lambda_n^{(1)}=\lambda_n$, and $\lambda_{n+i}^{(k)}$ are the eigenvalue of the matrix $\bR_{n+i}$ associated with the eigenvector which is the $[(n+i+k-2 \hspace{1mm}\mbox{mod}\hspace{1mm} M) + 1]$-th column of the Hadamard matrix ${\bf H}$ for all $k=1,2,\ldots,M$ and for all $i\in \{0,1\}$.
\end{proposition}
\begin{IEEEproof}
The proof of Proposition~\ref{ic:pro1} is given in Appendix~\ref{proofofpro1}.
\end{IEEEproof}
\Red{\begin{remark} \label{suprmk1} The use of modulated coefficients is a mathematical trick to force the covariance matrices $\bR_n$ among all transmitted signals to have a fixed set of eigenvectors at each time $n=1,2,\ldots$ Thanks to this forcing mechanism, a relation between the eigenvalues of $\bR_n$ and $\bR_{n+1}$ (and/or $\bR_n$ and $\bR_1$) can be established. This trick was first introduced by Ozarow and Leung for $2$-user Gaussian MAC~\cite{Ozarow} and for $2$-user Gaussian broadcast channel~\cite{Ozarow1984} in 1984. Kramer generalized this idea to design feedback codes based on Discrete Fourier Transform matrix (DFT) for $M$-user complex symbol IC channels~\cite{Kramer2002a} in 2002. Later, Truong and Yamamoto~\cite{TruongYamamoto2015e} combined these ideas with posterior matching idea~\cite{ShayevitzF} to design a feedback code for $2$-user IC which outperforms Kramer code~\cite{Kramer2002a} and Suh-Tse code for $2$-user IC real symbol channel~\cite{Suh2009e,Suh2011a}. They also designed a feedback code  which achieves the (optimal) linear feedback sum-capacity for the Gaussian broadcast channel~\cite{TruongYamamoto17a}.
\end{remark}}

\begin{proposition}
Every normalized covariance matrix ${\bf R}_n$ \red{has} all the columns of the $M\times M$ Hadamard matrix as its eigenvectors.
\end{proposition}
\begin{IEEEproof}
Applying Lemma~\ref{ic:lem1} with noting that $X \sim \mathcal{N}(0,P_1)$, we see that
\begin{align}
\bbE[(X_1^{(1)})^2]&= \bbE[F_X^{-1}(\Theta_1)F_X^{-1}(\Theta_1)]=P_1,\\
\bbE[(X_1^{(2)})^2]&= \bbE[F_X^{-1}(\Theta_2)F_X^{-1}(\Theta_2)]=P_1,\\
&\qquad \vdots \nonumber \\
\bbE[(X_1^{(M)})^2]&= \bbE[F_X^{-1}(\Theta_M)F_X^{-1}(\Theta_M)]=P_1.
\end{align}
Hence, we have
\begin{align}
\bbE[(X_1^{(1)})^2]=\bbE[(X_1^{(2)})^2]=\cdots=\bbE[(X_1^{(M)})^2]=P_1.
\end{align}
Besides, since $\Theta_m$ and $\Theta_k$ are pairwise independent for $m\neq k$, we also have
\begin{align}
\bbE[X_1^{(m)}X_1^{(k)}]=\bbE[F_X^{-1}(\Theta_m) F_X^{-1}(\Theta_k)]=0, \quad \forall m \neq k.
\end{align}
It follows that
\begin{align}
{\bf R}_1&=\frac{1}{P_1}\left[ \begin{array}{cccc}\bbE[X_1^{(1)} X_1^{(1)}] &\cdots&\bbE[X_1^{(1)}X_1^{(M)}]\\ \bbE[X_1^{(2)}X_1^{(1)}] &\cdots&\bbE[X_1^{(2)}X_1^{(M)}]\\ \vdots&\ddots&\vdots\\ \bbE[X_1^{(M)}X_1^{(1)}] &\cdots& \bbE[X_1^{(M)}X_1^{(M)}]\end{array}\right] \\
&={\bf I}_M,
\end{align}
where ${\bf I}_M$ is the $M\times M$ identity matrix.

By using the induction arguments and the fact that the indentity matrix ${\bf I}_M$ has all the columns of the Hadamard matrix ${\bf H}$ as its eigenvectors, together with the results of Propostion 1, we come to the conclusion. Note that we also have $\lambda_1^{(1)}=\lambda_1^{(2)}=\ldots=\lambda_1^{(M)}=1$.
\end{IEEEproof}

Now, we show that some other well-known coding schemes are special variants of our coding strategy above.
\subsection{Case of no interference $(a=0)$} \label{subsec51}
In this case,~\eqref{ic:eq62} and~\eqref{ic:eq64} in Appendix~\ref{proofofpro1}, which is the proof of  Proposition~\ref{ic:pro1}, become
\begin{align}
\label{ic:eq76}
P_{n+1}&=\frac{P_n}{\beta_n^2}\left[(1-b_n)^2+\frac{b_n^2}{P_n}\right],\\
\label{ic:eq77}
P_{n+1}{\bf R}_{n+1}&=\frac{P_n}{\beta_n^2}\left[(1-b_n)^2 {\bf R}_n +\frac{b_n^2}{P_n}{\bf I}_M\right].
\end{align}
By setting the pair ($P_n,b_n)$ as 
\begin{align}
\label{ic:eq78}
P_n &=P,\\
\label{ic:eq79}
b_n&=\frac{P}{P+1},
\end{align}
we obtain from~\eqref{ic:eq76} and~\eqref{ic:eq77} that
\begin{align}
\beta_n&=\frac{1}{\sqrt{P+1}},\\
{\bf R}_{n+1}&=\frac{1}{1+P}{\bf R}_n +\frac{P}{1+P}{\bf I}_M.
\end{align}
Since ${\bf R}_1={\bf I}_M$, we have ${\bf R}_n = {\bf I}_M$ for $n=1,2,...$.

In the non-interference case, the Gaussian interference channel with feedback becomes $M$ separate point-to-point Gaussian channels with feedback. Our coding algorithm with the parameters given by~\eqref{ic:eq78} and~\eqref{ic:eq79} coincides with Shayevitz and Feder's posterior matching scheme~\cite{ShayevitzF},~\cite{Shayevitz2007e} (or a variant of Schalkwijk-Kailath's scheme~\cite{SK66},~\cite{SK66Apr}). It is well-known that this coding scheme achieves the capacity of the channel.
\subsection{Case of two transmitter and two receivers $(M=2)$}\label{subsec52}
In the special case $M=2$, denote $\rho_n:=\rho_n^{(1,2)}$ for simplicity. It is easy to show that $\lambda_n=1+|\rho_n|$ and $\alpha_n^{(1)}\alpha_n^{(2)}=\mbox{sgn}(\rho_n)$. By substituting these relations into~\eqref{ic:eq63} in Appendix~\ref{proofofpro1}, we have
\begin{align}
&P_{n+1}\rho_{n+1} \nn\\
&\qquad =\frac{P_n}{\beta_n^2} \bigg([1-b_n(1-a)]^2\rho_n  + ab_n(1+|\rho_n|)\nn\\
&\qquad \qquad \times [2(1-a)b_n+2ab_n-2]\mbox{sgn}(\rho_n) \bigg) \nn \\
\label{ic:eq83}
&\qquad=\frac{P_n\mbox{sgn}(\rho_n)}{\beta_n^2}\bigg(|\rho_n| - 2 b_n(|\rho_n|+a) \nn\\
&\qquad \qquad + b_n^2[|\rho_n|(1+a^2)+2a]\bigg).
\end{align}
On the other hand, we can show from~ \eqref{ic:eq62} that
\begin{align}
\label{ic:eq84}
P_{n+1}&=\frac{1}{\beta_n^2}\bigg(P_n-2P_nb_n[1+a|\rho_n|\nn\\
&\qquad +b_n^2[1+P_n+a^2P_n+2a|\rho_n|P_n]\bigg).
\end{align}
\eqref{ic:eq83} and~\eqref{ic:eq84} coincide with the equations (6) and (7) in~\cite{TruongYamamoto2015e}. By setting 
\begin{align}
P_n&=P,\\
b_n&=\frac{P(1+a|\rho_n|)}{P(1+a^2+2a|\rho_n|)+1}
\end{align}
into~\eqref{ic:eq84}, we obtain
\begin{align}
\beta_n=\sqrt{\frac{a^2P(1-|\rho_n|^2)+1}{P(1+a^2+2a|\rho_n|)+1}}.
\end{align}
\red{With this choice of parameters, we obtain a new code which is an optimized version of Kramer code~\cite[Sec. VI-B]{Kramer2002a}. In Sections~\ref{2user} and~\ref{sec:num}, we will show that this variant code outperforms the Kramer code for all channel parameters. }   
\subsection{Case of $SNR=INR\enspace (a=1)$}\label{subsec53}
This special case has been considered in~\cite{Kramer2002a}. From~\eqref{ic:eq66}, we have
\small
\begin{align}
\label{ic:eq88}
P_{n+1}\lambda_{n+1}^{(k)}=\begin{cases}\frac{P_n}{\beta_n^2}\left[\lambda_n^{(k+1)}+\frac{b_n^2}{P_n}\right],&\hspace{2mm} k<M, \\\frac{P_n}{\beta_n^2}\left[\lambda_n^{(1)}+Mb_n\lambda_n^{(1)}(Mb_n-2)+\frac{b_n^2}{P_n}\right],&\hspace{2mm} k=M.\end{cases}
\end{align}
\normalsize
Choose
\begin{align}
\beta_n&=\sqrt{\frac{P\lambda_n(M-\lambda_n)+1}{PM\lambda_n+1}},\\
b_n&=\frac{P_n\lambda_n}{P_nM\lambda_n+1}.
\end{align}
Then, from~\eqref{ic:eq62}, we have $P_{n+1}=P_n$. Therefore, if we set $P_1=P$ as Kramer code~\cite{Kramer2002a}, then we have $P_n=P$ for $\forall n\in \mathbb{N}$, i.e., the input power constraint is satisfied. Besides, from the relation~\eqref{ic:eq88} we also have
\begin{align}
\label{ic:eq91}
\lambda_{n+1}^{(k)}=\begin{cases}\frac{(PM\lambda_n^{(1)}+1)\lambda_n^{(k+1)}-P(1-a_n)(\lambda_n^{(1)})^2}{P\lambda_n^{(1)}(M-\lambda_n^{(1)})+1},&\hspace{2mm} k < M,\\\frac{(PM\lambda_n^{(1)}+1)\lambda_n^{(1)}-P[1+(M-1)a_n](\lambda_n^{(1)})^2}{P\lambda_n^{(1)}(M-\lambda_n^{(1)})+1},&\hspace{2mm} k=M,\end{cases}
\end{align}
where 
\begin{align}
a_n = 1+\frac{1}{PM\lambda_n^{(1)}+1}.
\end{align}
The equation~\eqref{ic:eq91} coincides with the one given by (76) in~\cite{Kramer2002a}. It is shown in~\cite{Kramer2002a} that for large $M$, the sum-rate is approximately $(\log M)/2 + \log\log M$. This rate is about $\log \log M$ larger than the sum-rate capacity without feedback, which is $\log(1+PM)/2 \approx (\log M)/2$ (cf. \cite{Kramer2002a}).
\section{Main Results}\label{sec7}
\begin{theorem}\label{ic:thm2}
The symmetric rate
\begin{align}
\label{symrot}
R_{\mathrm{sym}}=\frac{1}{2}\hspace{1mm} \log^+ \frac{1}{\beta^2}
\end{align}
is achievable if the following relations hold for a triplet $(b,\beta,\lambda)$.
\begin{align}
\label{ic:eq94}
&\beta^2=[1-b(1-a)]^2 + ab\lambda[2(1-a)b + Mab-2]+\frac{b^2}{P}, \\
\label{ic:eq95}
&0< \lambda < M,\\
\label{ic:eq96}
&(A\neq 0, A\neq C, 
\frac{\lambda^{(k)}-\lambda^{(k+1)}}{A-C}>0, \hspace{1mm}
\forall k<M) \nonumber \\
&\hspace{24mm} \rm{or} \hspace{1mm}(A=C\neq 1, \lambda^{(k)}=\lambda, \hspace{1mm}\forall k\leq M),  \\
\label{ic:eq97}
&(1-CA^{M-1})\lambda=B(A^{M-1}+A^{M-2}+\cdots+A+1),
\end{align}
\red{where}
\begin{align}
\log^+(x)&:=\max\{\log_2 x, 0\},\\
\label{lamcond1}
\lambda^{(1)}&:=\lambda, \\
\label{ic:eq99}
\lambda^{(k+1)}&=\frac{\lambda^{(k)}-B}{A},\hspace{1mm}k=1,2,\ldots,M-1,\\
\label{ic:eq100}
A&=\frac{[1-b(1-a)]^2}{\beta^2},\\
\label{ic:eq101}
B&=\frac{b^2}{P\beta^2},\\
\label{ic:eq102}
C&=\frac{[1-b(1-a)]^2+Mab[2(1-a)b+Mab-2]}{\beta^2}.
\end{align}
\end{theorem}
\begin{IEEEproof}
The proof of Theorem~\ref{ic:thm2} is given in Appendix~\ref{proofofthm2}.
\end{IEEEproof}

\subsection{Case of no interference $(a=0)$}\label{subsec71}
\begin{corollary} Under no interference ($a=0$), the time-varying coding scheme can achieve the following symmetric rate
\begin{align}
R_{\mathrm{sym}}=\frac{1}{2}\log(P+1).
\end{align}
\end{corollary}
\begin{IEEEproof}
In the case of $a=0$, we see from~\eqref{ic:eq100}--\eqref{ic:eq102} and~\eqref{ic:eq94} that
\begin{align}
\label{ic:eq144}
A=C=\frac{(1-b)^2}{\beta^2},\quad B=\frac{b^2}{P\beta^2},
\end{align}
and
\begin{align}
\label{ic:eq145}
\beta^2 =(1-b)^2 + \frac{b^2}{P}.
\end{align}
From~\eqref{ic:eq111} in Appendix~\ref{proofofthm2},~\eqref{ic:eq144},~\eqref{ic:eq145}, $\lambda$ satisfies
\begin{align}
\lambda=\frac{b^2}{P\beta^2-P(1-b)^2}=1 \in (0, M). 
\end{align}
Furthermore, from~\eqref{ic:eq99},~\eqref{ic:eq144},~\eqref{ic:eq145} we also have that
\begin{align}
\lambda^{(k)}=1, \hspace{1mm} \forall k=1,2,...,M.
\end{align}
It is easy to see from~\eqref{ic:eq145} that the minimum of $\beta^2$ is equal to $1/(P+1)$ which is obtained by $b=P/(P+1)$. In this case, we have that
\begin{align}
A=C=\frac{1}{P+1}\neq 1.
\end{align}
Therefore, all the conditions in Theorem 2 are satisfied, and accordingly, the achievable symmetric rate is given by
\begin{align}
R_{\mathrm{sym}}=\frac{1}{2} \log^+ \frac{1}{\beta^2}= \frac{1}{2}\log(P+1).
\end{align}
It is well known that the above $R_{\mathrm{sym}}$ is the symmetric capacity of this channel.
\end{IEEEproof}
\subsection{Case of two transmitter and two receivers $(M=2)$}\label{2user}
\begin{corollary}\label{cor2cor2} For a given $a$ and $P$, let
\begin{align}
\label{xirate}
\xi(\rho,b):=\frac{P}{P+b^2[1+P + a^2 P + 2a P \rho] - 2 bP[1+a\rho]}.
\end{align}
Then, the non-degraded symmetric Gaussian interference channel ($a > 0$) can achieve the following symmetric rate $R_{\rm{sym}}$ (bits/channel use):
\begin{align}
\label{banana1}
R_{\rm{sym}} =\frac{1}{2}\max_{\rho \in [0,\rho_0], \enspace b \in \{b_1^*, b_2^*\}}\log^+\xi(\rho,b),
\end{align}
where 
\begin{align}
0< \rho_0: =\sqrt{\frac{a^2P^2 + P -\sqrt{P[2a^2P^2 +P]}}{a^2P^2}}<1,
\end{align}
and
\begin{align}
b^*_{1,2} &= \frac{2P\rho + aP + a P\rho^2}{2a P + 2P\rho + 2a^2 P \rho + \rho + 2a P \rho^2} \nn\\
&\qquad \pm \frac{\sqrt{P^2 a^2 \rho^4 - 2\rho^2(a^2P^2 +P)+a^2 P^2}}{2a P + 2P\rho + 2a^2 P \rho + \rho + 2a P \rho^2}.
\end{align}
\end{corollary}
\begin{IEEEproof}
In the case of $M=2$, the normalized correlation matrix is
\begin{align}
{\bf R}_n=\left[\begin{array}{cc}1&\rho_n\\\rho_n&1\end{array}\right],
\end{align}
and $2\times 2$ Hadamard matrix is
\begin{align}
{\bf H}=\left[\begin{array}{cc}1&1\\1&-1\end{array}\right].
\end{align}
It is easy to see that the two eigenvalues of ${\bf R}_n$ associated with two columns of ${\bf H}$ are $1\pm \rho_n$. Therefore, the assumption that $\lambda_n \rightarrow \lambda$ is equivalent to the assumption that $|\rho_n| \rightarrow \rho \in [0,1]$ for $\rho=\lambda-1$.

From~\eqref{ic:eq94}, we have
\begin{align}
\label{ic:eq156}
P=\frac{1}{\beta^2}[P-2bP(1+a\rho)+b^2(1+P+a^2P+2a\rho P)].
\end{align}
On the other hand, it also holds from~\eqref{ic:eq94},~\eqref{ic:eq100}, and~\eqref{ic:eq101} that
\begin{align}
\label{ic:eq157}
\frac{ab[2(1-a)b+Mab-2]}{\beta^2}=\frac{1-A-B}{\lambda}.
\end{align}
Therefore, we obtain from~\eqref{ic:eq100},~\eqref{ic:eq102},~\eqref{ic:eq157} that
\begin{align}
C=A+\frac{2}{\lambda}\left[1-(A+B)\right].
\end{align}
By substituting $M=2$ into~\eqref{ic:eq97}, we obtain
\begin{align}
(1-CA)\lambda=B(A+1).
\end{align}
This leads to
\begin{align}
\lambda&=CA\lambda + B(A+1) \nonumber \\
&=A\lambda\left(A+\frac{2}{\lambda}[1-(A+B)]\right)+B(A+1)  \nonumber \\
&=A^2 \lambda + 2A[1-(A+B)]+B(A+1)  \nonumber \\
&=A^2\lambda+2A(1-A)+B(1-A).
\end{align}
Equivalently,
\begin{align}
\label{ic:eq165}
\lambda(1-A^2)=(1-A)(2A+B).
\end{align}
The relation~\eqref{ic:eq165} holds if $\lambda$ satisfies $\lambda(1+A)=2A+B$, which means that
\begin{align}
(1+\rho)(1+A)=2A +B 
\end{align}
or
\begin{align}
\rho(1+A)=2A+B-1-A=A+B-1.
\end{align}
Then
\begin{align}
-\rho&=1-A-B+\rho A  \nonumber \\
&=1-\frac{[1-b(1-a)]^2}{\beta^2}-\frac{b^2}{P\beta^2}+\rho\frac{[1-b(1-a)]^2}{\beta^2}   \nonumber \\
&=\frac{1}{\beta^2}\left(\beta^2-[1-b(1-a)]^2 -\frac{b^2}{P}+\rho[1-b(1-a)]^2\right)  \nonumber \\
\label{ic:eq171}
&=\frac{1}{\beta^2}\bigg[\rho-2b(\rho+a)+b^2(\rho(1+a^2)+2a)\bigg],
\end{align}
where the last equality holds from~\eqref{ic:eq94}. Note that we assume that $a> 0$ in~\ref{subsec22}.

\red{Combining~\eqref{ic:eq156} with~\eqref{ic:eq171}, we come to an equation system with three unknowns $(b, \rho, \beta)$ as follows: 
\begin{align}
\label{eqsys0}
P &= \frac{1}{\beta^2} \bigg[ P - 2 b P (1 + a \rho) + b^2 (1 + P + a^2 P + 2 a \rho P)\bigg],\\
\label{eqsys1}
-\rho &= \frac{1}{\beta^2} \bigg[\rho - 2b(\rho +a) + b^2\big(\rho(1+a^2) + 2 a\big)  \bigg].
\end{align}}

By considering $\rho$ as a variable, we obtain the following quadratic equation in $b$ for each fixed choice of $\rho$:
\begin{align}
\label{eq142mod}
&b^2[2 a P + 2P\rho + 2 a^2 P \rho + \rho + 2 a P \rho^2]  \nn\\
&\quad - 2b[2P\rho + a P + P a\rho^2] + 2P\rho =0.
\end{align}
The discriminant of this quadratic equation is given by
\begin{align}
\Delta = P^2 a^2 \rho^4 - 2\rho^2(a^2 P^2 +P) +a^2 P^2: = f(\rho).
\end{align}
Since $f(0) = a^2 P^2 > 0 $ and $f(1) = -2P <0$, there exists the minimum value $\rho_0 \in (0,1)$ such that $f(\rho_0)=0$. Furthermore, it can be shown that the value of $\rho_0$ satisfies
\begin{align}
0< \rho_0 =\sqrt{\frac{a^2P^2 + P -\sqrt{P[2 a^2P^2 +P]}}{a^2P^2}}<1.
\end{align}
On the other hand, the first derivative satisfies
\begin{align}
f'(\rho) =  4P^2 a^2 (\rho^3-\rho) - 4P\rho \leq 0
\end{align} for all $\rho \in [0,1]$. This means $\Delta = f(\rho) \geq 0$  for all $\rho \in [0, \rho_0]$. For all these values of $\rho$, it can easily be shown that~\eqref{eq142mod} has two positive solutions $b^*_1, b^*_2$ described in Corollary 2. In short, the equation system~\eqref{ic:eq156} and~\eqref{ic:eq171} has at least one solution $(b,\beta)$ for each fixed  $\rho \in [0,\rho_0]$. From Theorem~\ref{ic:thm2},  the symmetric Gaussian interference channel with feedback can achieve the following rate 
\begin{align}
\label{R152eq}
R_{\mathrm{sym}}&= \frac{1}{2} \log^+\bigg(\frac{1}{\min_{\rho \in [0,\rho_0]} \beta^2}\bigg) \nn\\
&= \frac{1}{2}\max_{\rho \in [0,\rho_0], \enspace b \in \{b_1^*, b_2^*\}}\log^+\xi(\rho,b),
\end{align} where $\xi(\rho,b)$ is defined in~\eqref{xirate}.
\end{IEEEproof}
\red{\begin{corollary}\label{lemconff}
For $M=2$, the proposed time-varying code outperforms the Kramer code in~\cite{Kramer2002a} for all channel parameters.
\end{corollary}
\begin{IEEEproof}
A variant of Kramer code is formed by setting the triple $(b,\beta, \rho)$, a solution of equation system~\eqref{eqsys0} and \eqref{eqsys1}, as follows:
\begin{align}
b&= \frac{P(1+a\rho)}{P(1+a^2+2 a\rho)+1},\\
\beta &= \sqrt{\frac{a^2P(1-\rho^2)+1}{P(1+a^2+2 a\rho)+1}},
\end{align}
and $\rho$ is the unique solution in $(0,1)$ of the equation:
\begin{align}
\label{eq156sup}
&2 a^3P^2\rho^4 + a^2P\rho^3-4 a P(a^2P+1)\rho^2 \nn\\
&\qquad -(2a^2P+P+2)\rho + 2 a P(a^2P+1)=0.
\end{align}
(See also in \cite{Kramer2002a,Kramer2004a, Suh2009e}). Our proposed code can optimize $\rho$ as shown in~\eqref{R152eq} while the Kramer code must use the special $\rho$ given by the solution of~\eqref{eq156sup}.
Therefore, the proposed code can outperform the Kramer code.
\end{IEEEproof}
\begin{remark} Numerical evaluations shown in Figs.~\ref{fig:HighSNRNew} and~\ref{fig:LowSNRnew} in Section~\ref{sec:num} affirm that the achievable rate of our coding scheme in Corollary~\ref{cor2cor2} is not worse than existing codes~\cite{Kramer2002a}, \cite{Suh2011a} for all channel parameters. These figures show that our coding scheme achieves better performance than Suh-Tse code~\cite{Suh2011a} when $\alpha = \log INR/\log SNR$ is not very large. In addition, our code can obtain better symmetric rate than (or at least equal to) the Kramer code  for all channel parameters, and therefore it overcomes all the weak-points of the Suh-Tse coding scheme and narrows the capacity gap to the Suh-Tse outer bound~\cite{Suh2011a}. 
\end{remark}}
\subsection{A variant of Kramer's code for $a=1$}\label{subsec73}
For $a=1$, by considering $\lambda \in [0,M]$ as a variable in~\eqref{ic:eq94}, we obtain the following quadratic equation in $b$:
\begin{align}
\beta^2=1+b\lambda(Mb-2)+\frac{b^2}{P}:=f_{\lambda}(b).
\end{align}
Without considering other constraints, the function $f_{\lambda}(b)$ has the derivative 
\begin{align}
f_{\lambda}'(b)=2Mb\lambda-2\lambda +\frac{2b}{P}.
\end{align}
Note that $f_{\lambda}'(b)=0$ if $b=b^*:=P\lambda/(MP\lambda+1)$. For this $b=b^*$, we obtain
\begin{align}
\beta^2=f_{\lambda}(b^*)=\frac{P\lambda(M-\lambda)+1}{PM\lambda+1},
\end{align}
and hence
\begin{align}
\beta=\sqrt{\frac{P\lambda(M-\lambda)+1}{PM\lambda+1}}.
\end{align}
In this case, we have from~\eqref{ic:eq100}--\eqref{ic:eq102} that
\begin{align}
\label{ic:eq181}
A&=\frac{1}{\beta^2}=\frac{PM\lambda+1}{P\lambda(M-\lambda)+1},\\
\label{ic:eq182}
B&=\frac{b^2}{P\beta^2}=\frac{P\lambda^2}{[P\lambda(M-\lambda)+1][PM\lambda+1]},\\
\label{ic:eq183}
C&=\frac{1+Mb(Mb-2)}{\beta^2}=\frac{1}{[P\lambda(M-\lambda)+1][PM\lambda+1]}.
\end{align}
Since $A\neq 1$, substituting~\eqref{ic:eq181},~\eqref{ic:eq182}, and~\eqref{ic:eq183} into~\eqref{ic:eq97} we obtain
\begin{align}
\lambda &= CA^{M-1}\lambda+B\frac{A^M-1}{A-1} \nonumber \\
&=\frac{\lambda}{[P\lambda(M-\lambda)+1][PM\lambda+1]}\left(\frac{PM\lambda+1}{P\lambda(M-\lambda)+1}\right)^{M-1}\nn\\
&\qquad +\frac{P\lambda^2}{[P\lambda(M-\lambda)+1][PM\lambda+1]}\frac{\left(\frac{PM\lambda+1}{P\lambda(M-\lambda)+1}\right)^M-1}{\frac{PM\lambda+1}{P\lambda(M-\lambda)+1}-1} \nn \\
&=\frac{\lambda}{[P\lambda(M-\lambda)+1][PM\lambda+1]}\left(\frac{PM\lambda+1}{P\lambda(M-\lambda)+1}\right)^{M-1} \nn\\
&\qquad +\frac{1}{PM\lambda+1}\left\{\left(\frac{PM\lambda+1}{P\lambda(M-\lambda)+1}\right)^M-1\right\}.
\end{align}
Hence,
\begin{align}
\lambda&= -\frac{1}{PM\lambda+1}\nn\\
&\quad + \left(\frac{PM\lambda+1}{P\lambda(M-\lambda)+1}\right)^M\frac{1}{(PM\lambda+1)^2}(\lambda+PM\lambda+1).
\end{align}
It follows that
\begin{align}
\lambda+\frac{1}{PM\lambda+1}=\left(\frac{PM\lambda+1}{P\lambda(M-\lambda)+1}\right)^M\frac{\lambda+PM\lambda+1}{(PM\lambda+1)^2}.
\end{align}
Hence,
\begin{align}
\label{ic:eq189}
\left(\lambda+\frac{1}{PM\lambda+1}\right)\left(\frac{P\lambda(M-\lambda)+1}{PM\lambda+1}\right)^M=\frac{\lambda+PM\lambda+1}{(PM\lambda+1)^2}.
\end{align}
Then, we have
\begin{align}
&\left(\lambda+\frac{1}{PM\lambda+1}\right)\left(\frac{P\lambda(M-\lambda)+1}{PM\lambda+1}\right)^M\nn\\
&\qquad \qquad =\frac{\lambda}{(PM\lambda+1)^2}+\frac{1}{PM\lambda+1}.
\end{align}
Rewrite this equation as
\begin{align}
\label{ic:eq191}
\lambda=&(MP\lambda+1)^2\bigg\{\left(\lambda+\frac{1}{PM\lambda+1}\right)\left(\frac{P\lambda(M-\lambda)+1}{PM\lambda+1}\right)^M \nn\\
&\qquad  -\frac{1}{PM\lambda+1}\bigg\}.
\end{align}
Observe that the equation~\eqref{ic:eq191} coincides with the equation (92) in the paper~\cite{Kramer2002a}\footnote{There is a typo in equation (92) in~\cite{Kramer2002a}.}. Besides, we have from~\eqref{ic:eq189} that
\begin{align}
\frac{PM\lambda+\lambda+1}{PM\lambda^2+\lambda+1}=(PM\lambda+1)\left(\frac{PM(M-\lambda)+1}{PM\lambda+1}\right)^M,
\end{align}
or
\begin{align}
\label{ic:eq193}
&\frac{PM\lambda^2+\lambda^2+\lambda}{PM\lambda^2+\lambda+1}\nn\\
&\qquad =\lambda(PM\lambda+1) \exp_2\left[M\log\left(\frac{PM(M-\lambda)+1}{PM\lambda+1}\right)\right].
\end{align}
Note that the equation~\eqref{ic:eq193} coincides with the equation (93) in~\cite{Kramer2002a}. Moreover, observe that
\begin{align}
A&=\frac{PM\lambda+1}{P\lambda(M-\lambda)+1}>1,\\
B&=\frac{P\lambda}{PM\lambda+1}>0.
\end{align}
Hence,
\begin{align}
\lambda^{(k+1)}=\frac{\lambda^{(k)}-B}{A}<\lambda^{(k)}, k=1,2,...,M.
\end{align}
This means that the condition~\eqref{ic:eq96} in Theorem~\ref{ic:thm2} is also satisfied, and therefore the achievable symmetric rate is given by
\begin{align}
R_{\mathrm{sym}}=\frac{1}{2}\log^+\frac{1}{\beta^2}=\frac{1}{2}\log\left[\frac{PM\lambda+1}{P\lambda(M-\lambda)+1}\right].
\end{align}
For $M$ sufficiently large, it is also shown in~\cite{Kramer2002a} that the sum-rate is about $\lambda/2$, which is approximately $(\log M/2)+\log\log M$. This sum-rate is about $\log\log M$ larger than the sum-rate capacity without feedback, which is $\log(1+PM)/2\approx (\log M)/2$.
\subsection{Extension of the Kramer's code for $a\neq 1$}\label{subsec74}
In the same way as Section~\ref{subsec73}, we consider the minimum value of the following equation for each fixed value of $a$:
\begin{equation}
\beta^2=[1-b(1-a)]^2+ab\lambda[2(1-a)b+Mab-2]+\frac{b^2}{P}:=g_{\lambda}(b).
\end{equation}
The first derivative of $g_{\lambda}(b)$ is
\begin{align}
g'_{\lambda}(b)&=\left(2(1-a)^2+\frac{2}{P}+[4a(1-a)+2Ma^2]\lambda\right)b \nn\\
&\qquad  -(2a\lambda+2(1-a)).
\end{align}
If \red{we do not care about other restrictions}, the function $g_{\lambda}(b)$ attains the minimum value at
\begin{align}
b^*=\frac{a\lambda+(1-a)}{(1-a)^2+1/P+[2a(1-a)+Ma^2]\lambda}.
\end{align}
The minimum value of $\beta^2$ in this case is
\begin{align}
\beta^2=g_{\lambda}(b^*).
\end{align}\red{However, for the case $a\neq 1$, it is not easy to show the existence of a $\lambda>0$ such that $(b^*,\sqrt{g_{\lambda}(b^*)},\lambda)$ satisfies all the restrictions in Theorem~\ref{ic:thm2}. In~\cite{Kramer2002a}, the coding scheme for $a\neq 1$ and $M>2$ was also not proposed. The main difficulty is the overwhelming computation which happens when $a\neq 1$. It is also known that this method of choosing parameters is suboptimal at least for $M=2$ as mentioned in Section~\ref{2user} (cf. also~\cite{Suh2009e,Suh2011a}). More specifically, it is shown in~\cite{Suh2009e} that the Kramer code for $M=2$ does not achieve the optimal generalized degree of freedom of the interference channel with feedback. In the following subsections, we show that a judicious choice of parameters of the time-varying code can achieves the generalized degree of this channel not only for $M=2$ but also for $M> 2$. Our time-varying code, which achieves the optimal generalized degree of freedom, is proposed for any value of $a\notin \{0,1\}$ and for any $M \in \bbZ^+$ where the Hadamard matrix exists. }

\subsection{Generalized Degree of Freedom of the Time-Varying Coding Scheme}\label{subsec75}
In the following, we will characterize the achievable symmetric rate as the solution of a quartic equation. 
\begin{theorem}\label{ic:thm3} For $a\notin \{0,1\}$, the following symmetric rate $R_{\rm{sym}}$ (bits/channel use) is achievable for $M$-user symmetric Gaussian channel with feedback:
\begin{align}
R_{\rm{sym}}= \frac{1}{2}\log^+\bigg[\frac{1}{\inf_{\textit{A}>1} \beta^2(\textit{A,a,P})}\bigg],
\end{align}
where $\beta=\beta(A,a,P)$ is the smallest positive real number satisfying the following constraints:
\begin{align}
\label{ic:eq203}
Z_4\beta^4 + Z_3 \beta^3 +Z_2 \beta^2 + Z_1\beta + Z_0=0,
\end{align}
and
\begin{align}
\label{ic:eq204}
0< \frac{\beta^2(1-A)-b^2/P}{ab[2(1-a)b+Mab-2]}< M,
\end{align}
\red{where}
\begin{align}
b&=\frac{1-\beta\sqrt{A}}{1-a},\\
\label{defineZ4}
Z_4&=Y_0+\frac{AY_2}{(1-a)^2}+\frac{a[(M-2)a+2]}{P(1-a)^4}A^2,\\
\label{defineZ3}
Z_3&=- 4(\sqrt{A})^3\frac{a[(M-2)a+2]}{P(1-a)^4}+ \frac{2a}{P(1-a)^3}(\sqrt{A})^3 \nn\\
&\qquad - \frac{2Y_2\sqrt{A}}{(1-a)^2}- \frac{Y_1\sqrt{A}}{1-a},\\
\label{defineZ2}
Z_2&=\frac{a[(M-2)a+2]}{P(1-a)^4}6A -\frac{2a}{P(1-a)^3}3A  \nn\\
&\qquad +\frac{Y_2}{(1-a)^2}+\frac{Y_1}{(1-a)},\\
\label{defineZ1}
Z_1&=-\frac{a[(M-2)a+2]}{P(1-a)^4}4\sqrt{A}+ \frac{2a}{P(1-a)^3}3\sqrt{A},\\
\label{defineZ0}
Z_0&=\frac{a[(M-2)a+2]}{P(1-a)^4}-\frac{2a}{P(1-a)^3},\\
\label{defineY0}
Y_0&= -\frac{(A^M-1)(A-1)}{\left(-MA^{M-1}+\frac{A^M-1}{A-1}\right)},\\
\label{defineY1}
Y_1&=\frac{M(A-1)A^{M-1}}{\left(-MA^{M-1}+\frac{A^M-1}{A-1}\right)}2a,\\
\label{defineY2}
Y_2&= -\frac{(A^M-1)}{P\left(-MA^{M-1}+\frac{A^M-1}{A-1}\right)} \nn\\
&\qquad -\frac{(M(A-1)A^{M-1})}{\left(-MA^{M-1}+\frac{A^M-1}{A-1}\right)}a[(M-2)a+2].
\end{align}
\end{theorem}
\begin{IEEEproof}
The proof of Theorem~\ref{ic:thm3} is given in Appendix~\ref{proofofthm3}. 
\end{IEEEproof}
From Theorem~\ref{ic:thm3}, we obtain the following corollaries.

\begin{corollary} \label{ic:cor4} For $\alpha = \log INR/ \log SNR > 1$, the generalized degree of freedom of the proposed coding scheme is given by
\begin{align}
d(\alpha)= \frac{\alpha}{2}.
\end{align}
\end{corollary}
\begin{corollary}\label{ic:cor5} For $\alpha = \log INR/ \log SNR < 1$, the generalized degree of freedom of the proposed coding scheme is given by
\begin{align}
d(\alpha)= 1-\frac{\alpha}{2}.
\end{align}
\end{corollary}
The proofs of Corollaries~\ref{ic:cor4} and~\ref{ic:cor5} are given in Appendices~\ref{proofofcor4} and~\ref{proofofcor5}, respectively.
\section{Numerical Results} \label{sec:num}
\begin{figure}[t]
	\centering
		\includegraphics[width=0.5\textwidth]{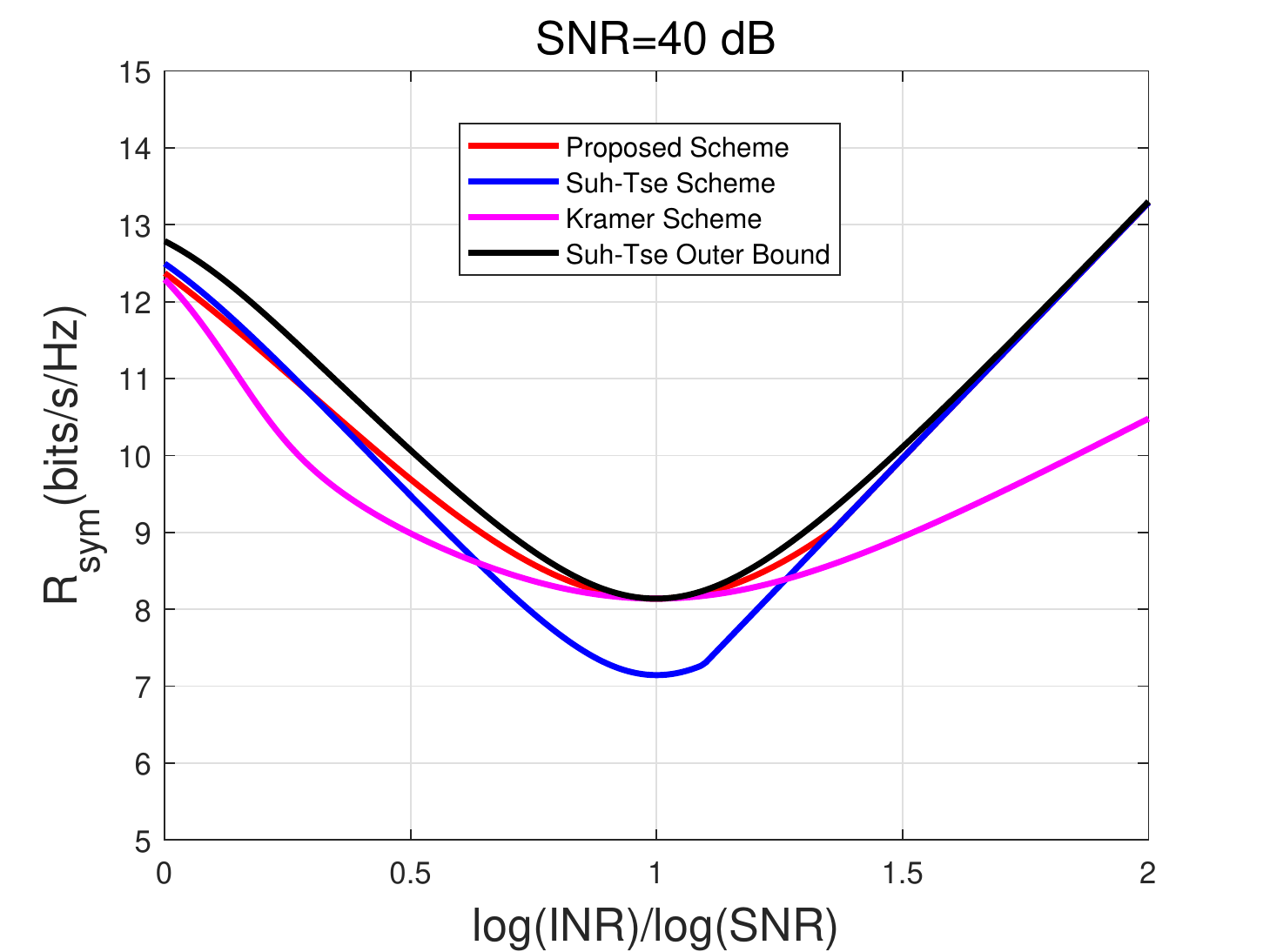}
	\caption{Symmetric rate comparison at high SNR}
	\label{fig:HighSNRNew}
\end{figure}

\begin{figure}[htbp]
	\centering
		\includegraphics[width=0.5\textwidth]{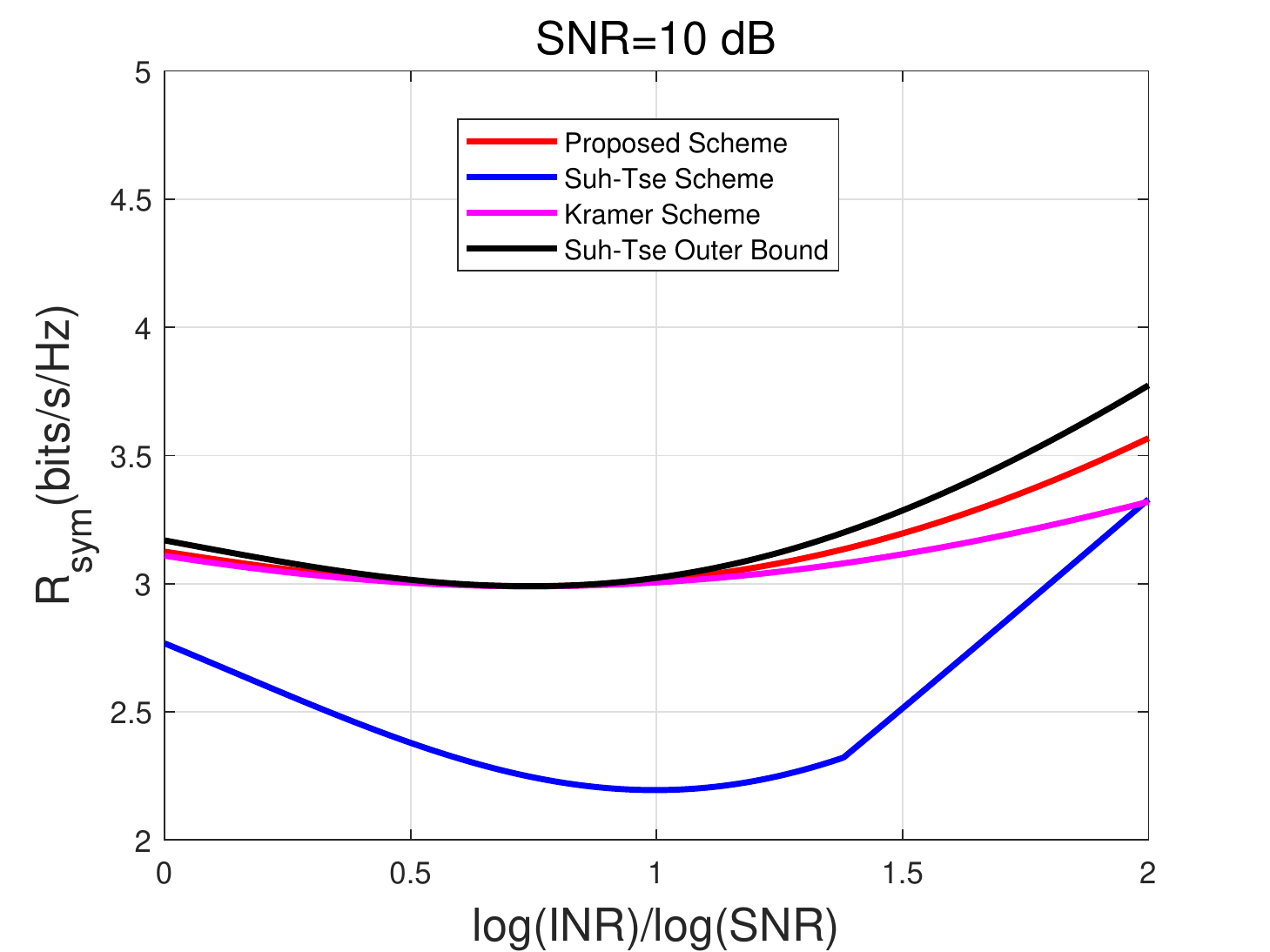}
	\caption{Symmetric rate comparision at low SNR}
	\label{fig:LowSNRnew}
\end{figure}

\begin{figure}[htbp]
	\centering
	\includegraphics[width=0.50\textwidth]{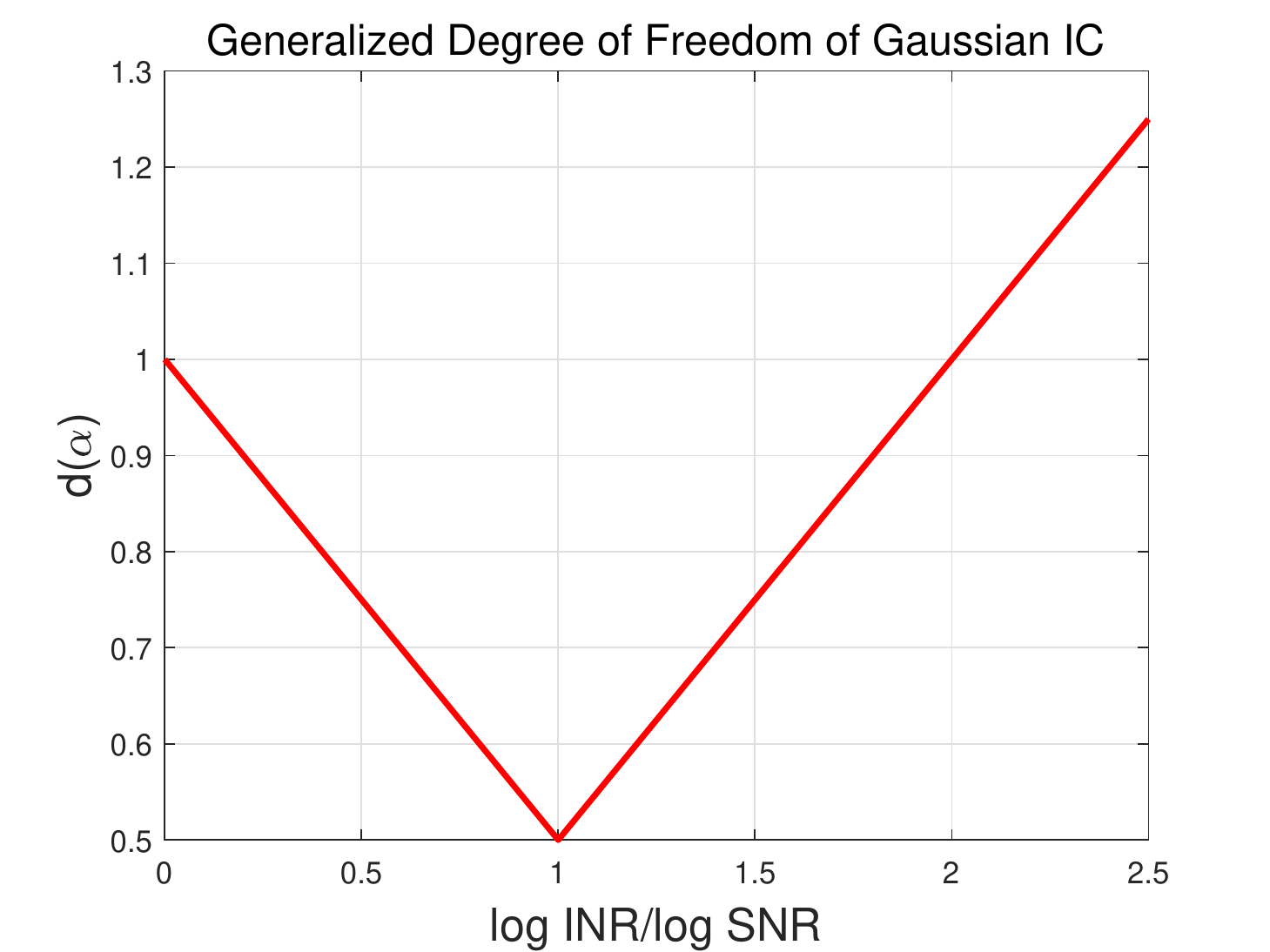}
	\caption{Generalized degree of freedom for feedback Gaussian IC}
	\label{fig:GDOF}
\end{figure}

\begin{figure}[htbp]
	\centering
		\includegraphics[width=0.50\textwidth]{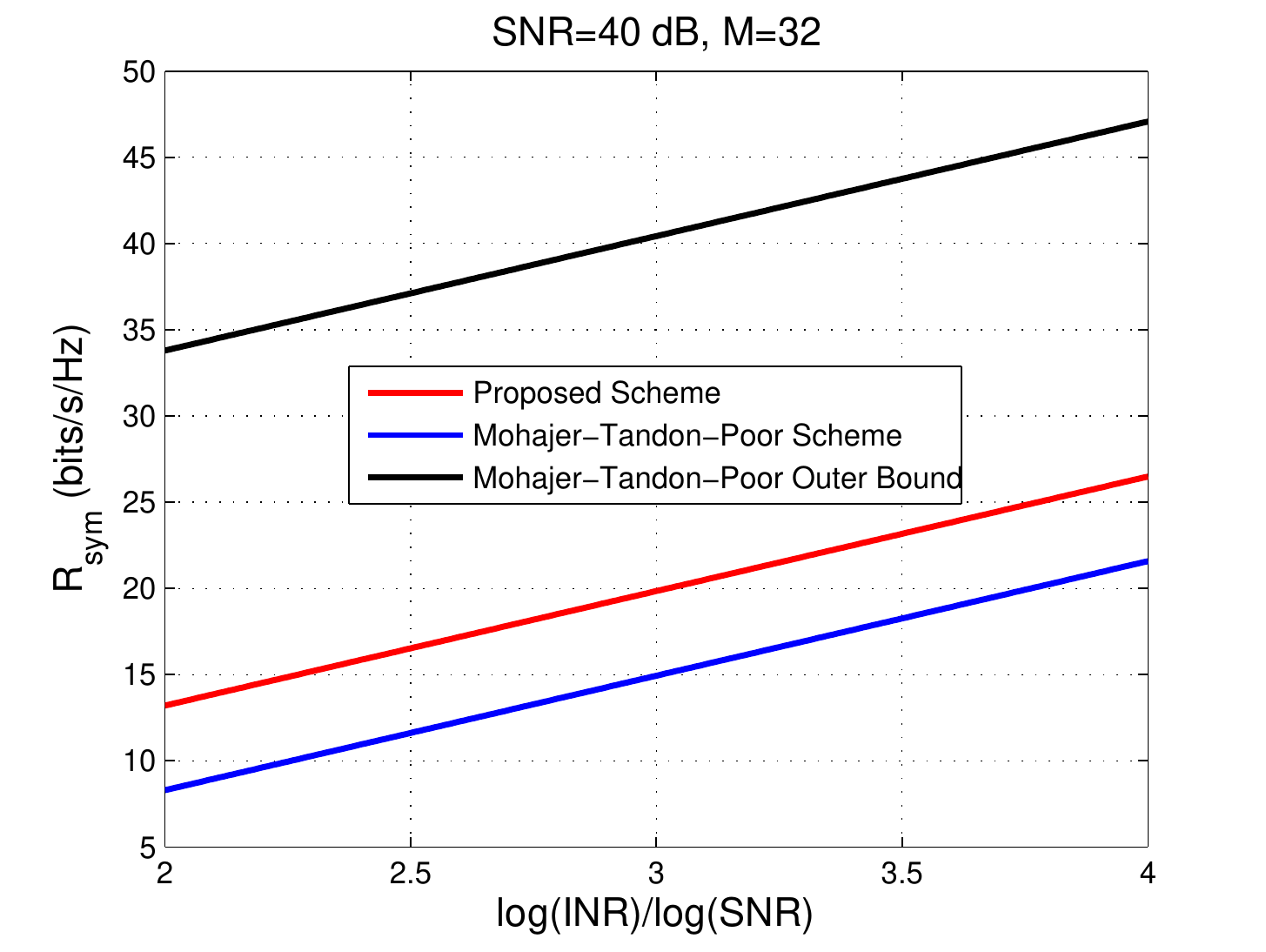}
	\caption{Symmetric rate comparison for the very strong Gaussian IC at high SNR}
	\label{fig:StrongIC}
\end{figure}

\begin{figure}[htbp]
	\centering
		\includegraphics[width=0.50\textwidth]{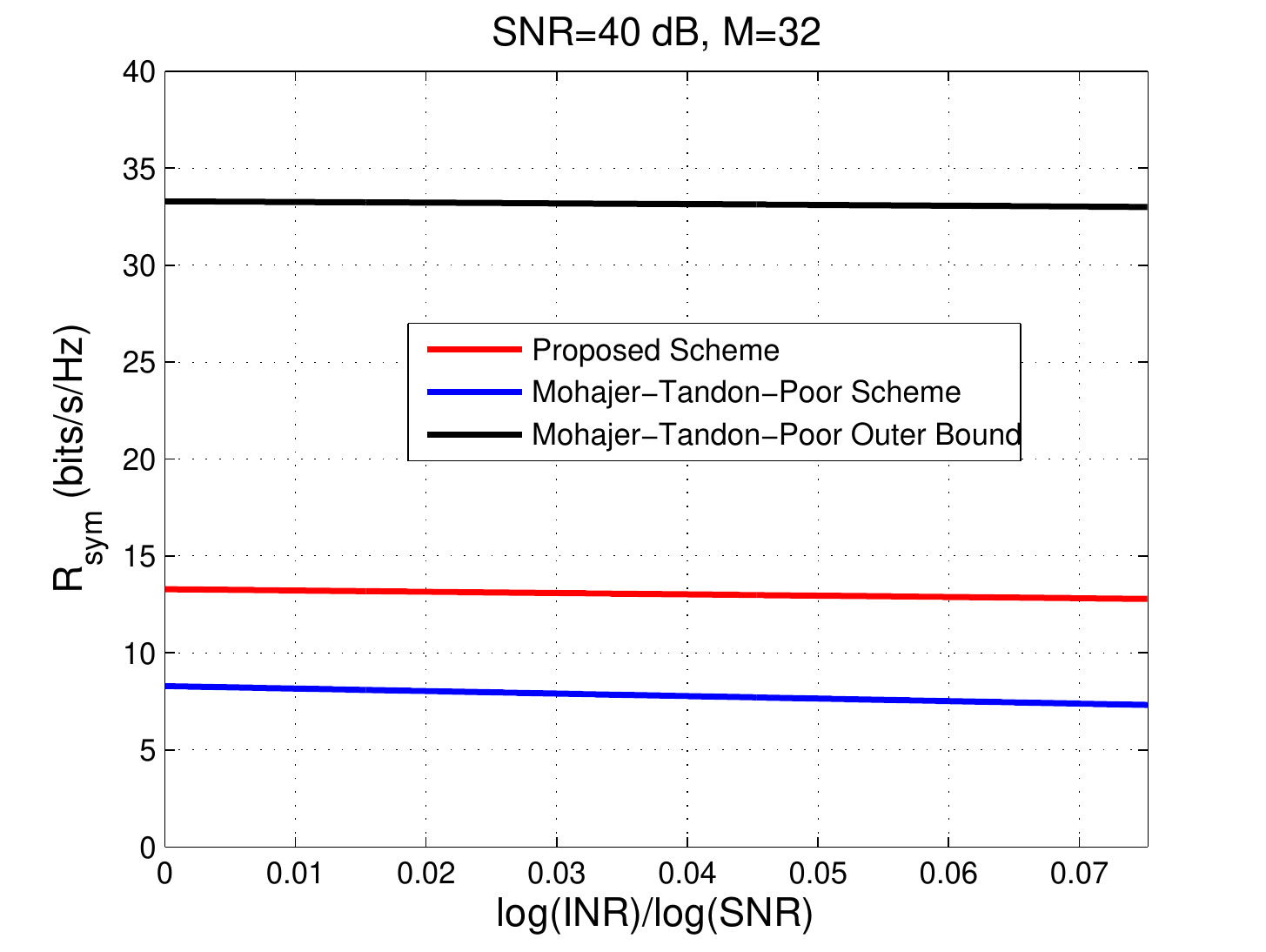}
	\caption{Symmetric rate comparison for the very weak Gaussian IC at high SNR}
	\label{fig:WeakIC}
\end{figure}

We have performed some numerical evaluations and obtains some results which affirm our mathematical arguments in this paper (cf. Figs.~\ref{fig:HighSNRNew}--~\ref{fig:WeakIC}). For the case $M=2$, in Figs.~\ref{fig:HighSNRNew} and~\ref{fig:LowSNRnew}, we show some numerical results of achievable symmetric rate for our proposed scheme in comparision with Suh-Tse scheme~\cite{Suh2011a}, Kramer scheme~\cite{Kramer2002a}, and Suh-Tse outer bound~\cite{Suh2011a}. These figures show that our coding scheme achieves better performance than Suh-Tse code when $\alpha = \log INR/\log SNR$ is not very large. In addition, our code can obtain better symmetric rate than (or at least equal to) the Kramer code for all channel parameters, and therefore it overcomes all the weak-points of the Suh-Tse coding scheme and narrows the capacity gap to the Suh-Tse outer bound. 

Fig.~\ref{fig:GDOF} draws the curve of the generalized degree of freedom of the fully-connected $M$-user Gaussian interference channel with feedback as a function of $\alpha = \log INR/ \log SNR$ for the case $\alpha\neq 1$ for any $M\geq 2$. This curve shows that the generalized degree of freedom $d(\alpha)$ is linearly decreasing and increasing in $\alpha<1$ and $\alpha>1$, respectively. For $\alpha=1$, the generalized degree of freedom  of this channel is not well-defined as shown in~\cite[Theorem 1]{Mohajer2013a}. This curve was shown to be optimal in~\cite[Theorem 1]{Mohajer2013a} for general $M$ or for the case $M=2$~\cite{Suh2011a}. Since other coding schemes~\cite{Suh2011a},~\cite{Mohajer2013a} which achieve the optimal generalized degree of freedom for the Gaussian interference channel with feedback are based on ``cooperative interference alignment", our results provide an important conclusion that the simple strategy ``treating other users as noise" also works well if interference channels allow feedback. Figs.~\ref{fig:StrongIC} and~\ref{fig:WeakIC} show that our coding scheme can even achieve better symmetric rate than the cooperative interference alignment strategy when numerically evaluated at some $M>2$.

\section{Conclusion}
A general time-varying posterior matching coding scheme for Gaussian interference channel with feedback was proposed. Based on the analysis of achievable symmetric rate of the channel, we proposed a coding scheme based on the ideas that a better cooperation strategy among transmitters will make the decoding process simpler and help to increase the achievable transmission symmetric rate.  All receivers only need to decode their intended messages. Our proposed code has been shown to narrow down the gap to the Suh-Tse outer bound for the case $M=2$. Besides, our code is optimal in the generalized degree of freedom sense for any $M\geq 2$. Our results show that the simple strategy ``treating other users as noise" also works well if interference channels allow feedback. An interesting future work is to find the way to characterize the achievable symmetric rate in a simpler way so that we can mathematically compare our code performance with other existing coding schemes. 
\appendices
\section{Proof of Proposition~\ref{ic:pro1}} \label{proofofpro1}
Define
\begin{align}
\delta(n)=\begin{cases}1,& n=0,\\0,&n\neq 0.\end{cases}
\end{align}
Firstly, observe that
\begin{align}
Y_n^{(m)} &= \alpha_n^{(m)} X_n^{(m)} + a\sum_{l=1, l \neq m}^M \alpha_n^{(l)} X_n^{(l)} + Z_n^{(m)}\\
          &= (1-a)\alpha_n^{(m)} X_n^{(m)} + a\sum_{l=1}^M \alpha_n^{(l)} X_n^{(l)} + Z_n^{(m)}.
\end{align}
By the transmission strategy, we have
\begin{align}
\label{sup42}
X_{n+1}^{(m)} &= \frac{1}{\beta_n} \left[X_n^{(m)}-b_n \alpha_n^{(m)} Y_n^{(m)}\right],\\
\label{sup43}
X_{n+1}^{(k)} &=\frac{1}{\beta_n} \left[X_n^{(k)}-b_n \alpha_n^{(k)} Y_n^{(k)}\right].
\end{align}
Hence, we obtain
\begin{align}
\label{ic:eq48}
&\bbE[X_{n+1}^{(m)} X_{n+1}^{(k)}]=\frac{1}{\beta_n^2}\bigg(\bbE[X_n^{(m)}X_n^{(k)}]-b_n\alpha_n^{(k)}\bbE[X_n^{(m)}Y_n^{(k)}]\nn\\
&\qquad -b_n \alpha_n^{(m)} \bbE[X_n^{(k)}Y_n^{(m)}] +b_n^2 \alpha_n^{(m)}\alpha_n^{(k)} \bbE[Y_n^{(m)} Y_n^{(k)}]\bigg).
\end{align}
Observe that
\begin{align}
&\bbE[X_n^{(m)} Y_n^{(k)}]=(1-a)P_n\alpha_n^{(k)}\rho_n^{(m,k)} + P_n a \sum_{l=1}^M \alpha_n^{(l)} \rho_n^{(m,l)},  \\
&\bbE[X_n^{(k)} Y_n^{(m)}]=(1-a)P_n\alpha_n^{(m)}\rho_n^{(k,m)} + P_n a \sum_{l=1}^M \alpha_n^{(l)} \rho_n^{(k,l)},\\
&\bbE[Y_n^{(m)} Y_n^{(k)}]\nn\\
&\qquad =\bbE\bigg([(1-a)\alpha_n^{(m)}X_n^{(m)}+a\sum_{l=1}^M \alpha_n^{(l)}X_n^{(l)} + Z_n^{(m)}] \nn\\
&\qquad \qquad \times [(1-a)\alpha_n^{(k)} X_n^{(k)} + a\sum_{t=1}^M \alpha_n^{(t)} X_n^{(t)} + Z_n^{(k)}]\bigg) \nn \\
&\qquad =(1-a)^2 P_n \alpha_n^{(m)} \alpha_n^{(k)}\rho_n^{(m,k)}\nn\\
&\qquad \qquad +P_na(1-a)\alpha_n^{(m)}\sum_{t=1}^M \alpha_n^{(t)}\rho_n^{(t,m)}\nn\\
&\qquad \qquad  +P_n a(1-a)\alpha_n^{(k)}\sum_{l=1}^M \alpha_n^{(l)}\rho_n^{(l,k)}\nn\\
&\qquad \qquad +a^2 P_n\sum_{l=1}^M\sum_{t=1}^M \alpha_n^{(l)}\alpha_n^{(t)} \rho_n^{(l,t)} + \delta(m-k).
\end{align}
Denote by $\lambda_n$ the eigenvalue associated with the eigenvector ${\boldsymbol \alpha}_n$ of the covariance matrix ${\bf R}_n$. 
Then, we have
\begin{align}
\sum_{l=1}^M\sum_{t=1}^M \alpha_n^{(l)}\alpha_n^{(t)}\rho_n^{(l,t)} &= {\boldsymbol \alpha}_n^T {\bf R}_n {\boldsymbol \alpha}_n = M\lambda_n,\\
\sum_{t=1}^M \alpha_n^{(t)} \rho_n^{(t,m)} &=  {\boldsymbol \alpha}_n^T{\boldsymbol \rho}^{(m)}_{n} = \alpha_n^{(m)}\lambda_n.
\end{align}
Therefore, using $\big(\alpha_n^{(m)}\big)^2=1$,
\begin{align}
\label{ic:eq55}
\bbE[X_n^{(m)} Y_n^{(k)}]&=(1-a)P_n\alpha_n^{(k)}\rho_n^{(m,k)} + P_n a \lambda_n \alpha_n^{(m)},\\
\label{ic:eq56}
\bbE[X_n^{(k)} Y_n^{(m)}]&=(1-a)P_n\alpha_n^{(m)}\rho_n^{(k,m)} + P_n a \lambda_n \alpha_n^{(k)},\\
\label{ic:eq57}
\bbE[Y_n^{(m)} Y_n^{(k)}]&=\delta(m-k)+ (1-a)^2P_n\alpha_n^{(m)}\alpha_n^{(k)}\rho_n^{(m,k)}\nn\\
&\qquad + 2P_n a(1-a)\lambda_n + a^2P M \lambda_n .
\end{align}
Substituting~\eqref{ic:eq55}--\eqref{ic:eq57} into~\eqref{ic:eq48}, we obtain
\begin{align}
&\bbE[X_{n+1}^{(m)} X_{n+1}^{(k)}]=\frac{1}{\beta_n^2}\bigg(P_n\rho_n^{(m,k)}\nn\\
& \qquad-b_n\alpha_n^{(k)}[(1-a)P_n\alpha_n^{(k)}\rho_n^{(m,k)} +P_n a\lambda_n\alpha_n^{(m)}]\nn\\
& \qquad -b_n\alpha_n^{(m)}[(1-a)P_n\alpha_n^{(m)}\rho_n^{(k,m)}+P_na\lambda_n\alpha_n^{(k)}]\nn \\
&\qquad  +b_n^2\alpha_n^{(m)}\alpha_n^{(k)}[\delta(m-k) +2P_na(1-a)\lambda_n\nn\\
& \qquad +a^2P_n M\lambda_n + (1-a)^2P_n \alpha_n^{(m)}\alpha_n^{(k)}\rho_n^{(m,k)}]\bigg) \nn \\
\label{ic:eq59}
&=\frac{P_n}{\beta_n^2}\bigg([1-b_n(1-a)]^2\rho_n^{(m,k)} \nn \\
&\qquad + a b_n\lambda_n[2(1-a)b_n+M a b_n-2]\alpha_n^{(m)}\alpha_n^{(k)} \nn\\
&\qquad + \frac{b_n^2}{P_n}\delta(m-k)\alpha_n^{(m)}\alpha_n^{(k)}\bigg).
\end{align}
By setting $k=m$ in~\eqref{ic:eq59}, we also obtain 
\begin{align}
\bbE[(X_{n+1}^{(m)})^2]&=\frac{P_n}{\beta_n^2}\bigg([1-b_n(1-a)]^2\nn\\
&\qquad + ab_n\lambda_n[2(1-a)b_n+Mab_n-2] +\frac{b_n^2}{P_n}\bigg),
\end{align}
for all $k=1,2,\ldots,M$.
This means that 
\begin{align}
\label{ic:eq61}
\bbE[(X_{n+1}^{(1)})^2]=\bbE[(X_{n+1}^{(2)})^2]=\cdots=\bbE[(X_{n+1}^{(M)})^2]:=P_{n+1}.
\end{align}
Therefore,
\begin{align}
\label{ic:eq62}
P_{n+1}&=\frac{P_n}{\beta_n^2}\bigg(\frac{b_n^2}{P_n}+[1-b_n(1-a)]^2 \nn\\
&\qquad + ab_n\lambda_n[2(1-a)b_n+Mab_n-2]\bigg) 
\end{align} for all $m=1,2,\ldots,M$.

On the other hand, from~\eqref{ic:eq41} and~\eqref{ic:eq59}, we obtain 
\begin{align}
\label{ic:eq63}
P_{n+1}\rho_{n+1}^{(m,k)}&=\frac{P_n}{\beta_n^2}\bigg([1-b_n(1-a)]^2\rho_n^{(m,k)}\nn\\
&\qquad + a b_n\lambda_n[2(1-a)b_n+M a b_n-2]\alpha_n^{(m)}\alpha_n^{(k)} \nn\\
&\qquad + \frac{b_n^2}{P_n}\delta(m-k)\alpha_n^{(m)}\alpha_n^{(k)}\bigg),
\end{align}
which means from~\eqref{ic2018:eq42} that
\begin{align}
\label{ic:eq64}
&P_{n+1}{\bf R}_{n+1}\nn\\
&\qquad =\frac{P_n}{\beta_n^2}\bigg([1-b_n(1-a)]^2{\bf R}_n+ \frac{b_n^2}{P_n}{\bf I}_M \nn\\
&\qquad \qquad  +ab_n\lambda_n[2(1-a)b_n + M a b_n -2]{\boldsymbol\alpha}_n{\boldsymbol \alpha}_n^T \bigg).
\end{align}

\Red{Let ${\boldsymbol \alpha}_n$ be the $[(n-1 \mod M)+1]$-th column of the Hadamard matrix ${\bf H}$ of order $M$, which is defined in Section~\ref{subsec21}. Recall the definition of $\bH_n$ in~\eqref{defHn}, where ${\bf H}_n=[\begin{array}{cccc}{\boldsymbol \alpha}_n & {\boldsymbol \alpha}_{n+1} & \ldots& {\boldsymbol \alpha}_{n+M-1}\end{array}]$.}

Then from~\eqref{ic:eq64}, we have
\begin{align}
\label{ic:eq65}
&P_{n+1}{\bf H}_{n+1}^T {\bf R}_{n+1}{\bf H}_{n+1}\nn\\
&=\frac{P_n}{\beta_n^2}\bigg([1-b_n(1-a)]^2{\bf H}_{n+1}^T{\bf R}_n {\bf H}_{n+1} + \frac{b_n^2}{P_n}{\bf H}_{n+1}^T{\bf H}_{n+1}\nn \\
&\qquad  +  ab_n\lambda_n[2(1-a)b_n + M a b_n -2]{\bf H}_{n+1}^T{\boldsymbol \alpha}_n{\boldsymbol \alpha}_n^T{\bf H}_{n+1}\bigg).
\end{align}
\red{Observe that
\begin{align}
&{\boldsymbol \alpha}_n^T {\bf H}_{n+1}\nn\\
&={\boldsymbol \alpha}_n^T\left[\begin{array}{ccccc}{\boldsymbol \alpha}_{n+1}&{\boldsymbol \alpha}_{n+2}&\ldots&{\boldsymbol \alpha}_{n+M-1}&{\boldsymbol \alpha}_{n+M}\end{array}\right] \nonumber \\
&=\left[\begin{array}{ccccc}0&0&\ldots&0&M\end{array}\right],
\end{align}
hence
\begin{align}
\label{sup62}
{\bf H}_{n+1}^T{\boldsymbol \alpha}_n{\boldsymbol \alpha}_n^T{\bf H}_{n+1}=\diag(0,0,\ldots,0,M^2).
\end{align}
In addition, we also have
\begin{align}
\label{eq63sup1}
{\bf H}_{n+1}^T{\bf H}_{n+1}&=M \bI_M
\end{align}
by the definition of the Hadamard matrix. Now, since $\bR_n$ has all the columns of the Hadamard matrix as its eigenvectors, it also has all the columns of ${\bf H}_{n+1}$ as its eigenvectors. Furthermore, since $\lambda_n^{(k)}$ is the eigenvalue of the matrix $\bR_n$ associated with the eigenvector which is the $[(n+k-2 \hspace{1mm}\mbox{mod}\hspace{1mm} M) + 1]$-th column of the Hadamard matrix ${\bf H}$ for all $k=1,2,\ldots,M$, by the EVD (Eigenvalue Decompostion), we have
\begin{align}
{\bf R}_n &={\bf H}_{n+1}\diag(\lambda_n^{(2)}, \lambda_n^{(3)},\ldots,\lambda_n^{(M)},\lambda_n^{(1)}){\bf H}_{n+1}^{-1} \nonumber \\
\label{eq64sup1}
&=\frac{1}{M} {\bf H}_{n+1}\diag(\lambda_n^{(2)}, \lambda_n^{(3)},\ldots,\lambda_n^{(M)},\lambda_n^{(1)}){\bf H}_{n+1}^T,
\end{align} where~\eqref{eq64sup1} follows from~\eqref{eq63sup1}. Furthermore, it follows from~\eqref{eq64sup1} that
\begin{align}
&{\bf H}_{n+1}^T{\bf R}_n {\bf H}_{n+1}\nn\\
&=\frac{1}{M} {\bf H}_{n+1}^T {\bf H}_{n+1}\diag(\lambda_n^{(2)}, \lambda_n^{(3)},\ldots,\lambda_n^{(M)},\lambda_n^{(1)}){\bf H}_{n+1}^T{\bf H}_{n+1} \nonumber \\
&=\frac{1}{M} M \bI_M \diag(\lambda_n^{(2)}, \lambda_n^{(3)},\ldots,\lambda_n^{(M)},\lambda_n^{(1)})M \bI_M \nonumber \\
\label{sup64}
&=M \diag(\lambda_n^{(2)}, \lambda_n^{(3)},\ldots,\lambda_n^{(M)},\lambda_n^{(1)}).
\end{align}
From~\eqref{ic:eq65},~\eqref{sup62},~\eqref{eq63sup1}, and~\eqref{sup64}, we have
\begin{align}
\label{ic:eq65sup}
&P_{n+1}{\bf H}_{n+1}^T {\bf R}_{n+1}{\bf H}_{n+1}\nn\\
&=\frac{P_n}{\beta_n^2}\bigg(\frac{b_n^2}{P_n}M \bI_M \nn\\
&+[1-b_n(1-a)]^2 M \diag(\lambda_n^{(2)}, \lambda_n^{(3)},\ldots,\lambda_n^{(M)},\lambda_n^{(1)})\nn\\
&+  ab_n\lambda_n[2(1-a)b_n + M a b_n -2]\diag(0,0,\ldots,0,M^2)\bigg).
\end{align}
As a result, the right hand side of the equation~\eqref{ic:eq65sup} is a diagonal matrix. In addition, from~\eqref{ic:eq65sup}, we also have
\begin{align}
&P_{n+1}{\bf R}_{n+1}{\bf H}_{n+1}\nn\\
&=\frac{1}{M}{\bf H}_{n+1} \frac{P_n}{\beta_n^2}\bigg([1-b_n(1-a)]^2 M \nn\\
& \qquad \times \diag(\lambda_n^{(2)}, \lambda_n^{(3)},\ldots,\lambda_n^{(M)},\lambda_n^{(1)})+ \frac{b_n^2}{P_n}M \bI_M\nonumber \\
& +  ab_n\lambda_n[2(1-a)b_n + M a b_n -2]\diag(0,0,\ldots,0,M^2)\bigg) \nonumber\\
\label{eq71:icsup}
&={\bf H}_{n+1}\frac{P_n}{\beta_n^2}\bigg([1-b_n(1-a)]^2 \nn\\
&\qquad \times  \diag(\lambda_n^{(2)}, \lambda_n^{(3)},\ldots,\lambda_n^{(M)},\lambda_n^{(1)})+ \frac{b_n^2}{P_n} \bI_M\nonumber \\
& +   ab_n\lambda_n[2(1-a)b_n + M a b_n -2]\diag(0,0,\ldots,0,M)\bigg).
\end{align}
From the relation~\eqref{eq71:icsup}, it is easy to see that all the columns of the matrix ${\bf H}_{n+1}$ are eigenvectors of $\bR_{n+1}$, so all the columns of the Hadamard matrix $\bH$. Furthermore, since $\lambda_{n+1}^{(k)}$ is the eigenvalue of the matrix $\bR_{n+1}$ associated with the eigenvector which is the $[(n+k-1 \hspace{1mm}\mbox{mod}\hspace{1mm} M) + 1]$-th column of the Hadamard matrix ${\bf H}$ for all $k=1,2,\ldots,M$, it holds that ${\bf R}_{n+1}{\bf H}_{n+1}={\bf H}_{n+1}\diag(\lambda_{n+1}^{(1)},\lambda_{n+1}^{(2)},\ldots,\lambda_{n+1}^{(M)})$. Hence, by~\eqref{eq71:icsup},~\eqref{ic:eq66} of Proposition~\ref{ic:pro1} holds. }
\section{Proof of Theorem~\ref{ic:thm2}}\label{proofofthm2}
Firstly, it is easy to see that if we can force $P_n \to P$, $\lambda_n^{(k)} \to \lambda^{(k)}, b_n \to b, \beta_n\to \beta$ for some triplet $(b,\beta,\lambda)$, then from~\eqref{ic:eq66} we have
\begin{align}
\lambda^{(k)}=\left\{\begin{array}{cc}A\lambda^{(k+1)}+B,& \hspace{2mm} k<M,\\C\lambda^{(1)} +B,&\hspace{2mm} k=M.\end{array}\right.
\end{align}
Therefore, we obtain
\begin{align}
\lambda &=\lambda^{(1)} \nonumber \\
&=A\lambda^{(2)}+B \nonumber  \\
&=A[A\lambda^{(3)}+B]+B \nonumber  \\
&=A^2\lambda^{(3)}+AB +B \nonumber \\
&\qquad \vdots \nonumber \\
&=A^{M-1}\lambda^{(M)}+A^{M-2}B+A^{M-3}B+\cdots+AB+B \nonumber  \\
&=CA^{M-1}\lambda + B(A^{M-1}+A^{M-2}+\cdots+A+1),
\end{align}
which means that
\begin{align}
(1-CA^{M-1})\lambda=B(A^{M-1}+A^{M-2}+\cdots+A+1),
\end{align}
and for the case $A\neq 1$, 
\begin{align}
\label{ic:eq111}
\lambda=\frac{B(A^M-1)}{(A-1)(1-CA^{M-1})}.
\end{align}
Moreover, the relation~\eqref{ic:eq94} holds from~\eqref{ic:eq62}.
We also note from~\eqref{ic:eq66} that all the other eigenvalues $\lambda^{(k)}$ satisfy the following relation:
\begin{align}
\label{eq100sup}
\lambda^{(k+1)}=\frac{\lambda^{(k)}-B}{A}, \hspace{1mm} k=1,2,\ldots,M-1.
\end{align}
(Note that, $\lambda^{(1)}=\lambda$). \\

In the next part, we show a procedure to realize $P_M=P$ and $\lambda_M^{(k)}=\lambda^{(k)}$ for all $k=1,2,\ldots,M$ by judiciously varying the values of $P_k, b_k, \beta_k$ for all $k=1,2,\ldots,M-1$.

Define
\begin{align}
\label{ic:eq114}
A_n&:=\frac{P_n[1-b_n(1-a)]^2}{P_{n+1}\beta_n^2},\\
\label{ic:eq115}
B_n&:=\frac{b_n^2}{P_{n+1}\beta_n^2},\\
\label{ic:eq116}
C_n&:=\frac{P_n([1-b_n(1-a)]^2+M a b_n[2(1-a)b_n + Mab_n -2])}{P_{n+1}\beta_n^2}.
\end{align}
From these definitions and~\eqref{ic:eq66} we have
\begin{align}
\label{ic:eq117}
\lambda_{n+1}^{(k)}=\left\{\begin{array}{cc} A_n\lambda_n^{(k+1)}+B_n,&\hspace{2mm} k<M,\\C_n\lambda_n^{(1)}+B_n,&\hspace{2mm} k=M.\end{array}\right.
\end{align}
Using the relation~\eqref{ic:eq117} recursively for $\Red{k=1,2,\ldots,M-1}$, we obtain that
\begin{align}
\lambda_M^{(k)}&=A_{M-1}\lambda_{M-1}^{(k+1)} + B_{M-1} \nonumber \\
&=A_{M-1}A_{M-2}\lambda_{M-2}^{(k+2)} \nn\\
&\qquad +A_{M-1}B_{M-2} + B_{M-1} \nonumber  \\
&\vdots \nonumber  \\
&= A_{M-1}A_{M-2}\ldots A_k \lambda_k^{(M)} \nn\\
&\qquad + A_{M-1} A_{M-2}\ldots A_{k+1}B_k \nn\\
&\qquad + A_{M-1} A_{M-2} \ldots A_{k+2}B_{k+1} \nn\\
&\qquad +A_{M-1}B_{M-2} + B_{M-1} \nonumber  \\
&=A_{M-1}A_{M-2}\ldots A_k C_{k-1}\lambda_{k-1}^{(1)} \nn\\
&\qquad + A_{M-1}A_{M-2}\ldots A_k B_{k-1}\nn\\
&\qquad + A_{M-1} A_{M-2}\ldots A_{k+1}B_k \nonumber\\
&\qquad+ A_{M-1} A_{M-2}\ldots A_{k+2}B_{k+1}\nn\\
&\qquad + A_{M-1}B_{M-2} + B_{M-1} \nonumber  \\
&=A_{M-1}A_{M-2}\ldots A_k C_{k-1}A_{k-2}\lambda_{k-2}^{(2)} \nn\\
&\qquad + A_{M-1}A_{M-2}\ldots A_k C_{k-1}B_{k-2} \nn\\
&\qquad  +A_{M-1}A_{M-2}\ldots A_k B_{k-1} \nonumber\\
&\qquad + A_{M-1} A_{M-2}\ldots A_{k+1}B_k  \nn\\
&\qquad + A_{M-1} A_{M-2}\ldots A_{k+2}B_{k+1}  \nn\\
&\qquad +A_{M-1}B_{M-2} + B_{M-1} \nonumber \\
&\vdots \nonumber \\
&=A_{M-1}A_{M-2}\ldots A_k C_{k-1} A_{k-2}\ldots A_1 \lambda_1^{(k-1)} \nn\\
&\qquad + A_{M-1} A_{M-2}\ldots A_k C_{k-1}A_{k-2}\ldots A_2B_1 \nonumber\\
&\qquad+ A_{M-1} A_{M-2}\ldots A_k C_{k-1}A_{k-2}\ldots A_3 B_2 \nn\\
&\qquad +A_{M-1}A_{M-2}\ldots A_k C_{k-1}B_{k-2}\nonumber\\
&\qquad + A_{M-1}A_{M-2}\ldots A_k B_{k-1} \nn\\
&\qquad +A_{M-1}B_{M-2} + B_{M-1} \nonumber  \\
\label{ic:eq124}
&\stackrel{(a)}{=}A_{M-1}A_{M-2}\ldots A_k C_{k-1} A_{k-2}\ldots A_1\nn\\
&\qquad + A_{M-1} A_{M-2}\ldots A_k C_{k-1}A_{k-2}\ldots A_2B_1\nonumber\\
&\qquad + A_{M-1} A_{M-2}\ldots A_k C_{k-1}A_{k-2}\ldots A_3 B_2 \nn\\
&\qquad +A_{M-1}A_{M-2}\ldots A_k C_{k-1}B_{k-2}\nonumber\\
&\qquad+ A_{M-1}A_{M-2}\ldots A_k B_{k-1}\nn\\
&\qquad +A_{M-1}B_{M-2} + B_{M-1}
\end{align}
for all $k=\Red{1,2,3,\ldots,M-1}$. Here (a) follows from the fact that ${\bf R}_1={\bf I}_M$, so $\lambda_1^{(k)}=1, \hspace{1mm} k=1,2,\ldots,M$.\\

We first fix $A_n$ and $C_n$ as $A_n=A$ and $C_n =C$. Then, we obtain the following relation from~\eqref{ic:eq124}:
\begin{align}
\lambda_M^{(k)}&=A^{M-2}C + A^{M-3}C B_1 + A^{M-4} C B_2 + \cdots \nn\\
&\qquad + A^{M-k}C B_{k-2} + A^{M-k}B_{k-1}\nn\\
&\qquad + A^{M-k-1}B_k+\cdots+AB_{M-2}+B_{M-1} \nonumber  \\
&=[A^{M-2}C + A^{M-3}C B_1 +\cdots\nn\\ 
&\qquad + A^{M-k}C B_{k-2}+ A^{M-(k+1)} C B_{(k+1)-2}]\nn\\
&\qquad -A^{M-(k+1)} C B_{(k+1)-2}+ A^{M-k} B_{k-1} \nonumber \\
&\qquad + [A^{M-(k+1)}B_k+\cdots+AB_{M-2}+B_{M-1}]  \nonumber \\
\label{ic:eq127}
&=\lambda_M^{(k+1)}-A^{M-k-1}C B_{k-1} + A^{M-k}B_{k-1}.
\end{align}
If $A=C \neq 1$, we have from~\eqref{ic:eq111} that
\begin{align}
\label{ic:eq128}
\lambda=\frac{B}{1-A}.
\end{align}
In addition, from~\eqref{ic:eq94},~\eqref{ic:eq101},~\eqref{ic:eq102}, and $A=C$,
\begin{align}
A&=C \nonumber \\
&=\frac{\beta^2-(b^2/P)}{\beta^2} \nonumber \\
&=1-\frac{b^2}{\beta^2P} \nonumber \\
\label{ic:eq130}
&=1-B.
\end{align}
This leads to $\lambda=1$. Note from~\eqref{ic:eq127} and $A=C$ that $\lambda_M^{(k)}=\lambda_M^{(k+1)},  k=1,2,\ldots,M-1$. Hence, by setting $P_k=P, \beta_k=\beta, b_k=b$ for all $k=1,2,\ldots,M-1$, we can achieve $\lambda_M^{(k)}=1=\lambda$ for all $k=1,2,\ldots,M$ and $P_M=P$ (from the relations~\eqref{ic:eq62} and~\eqref{ic:eq94}).

If $A \neq 0, A\neq C$, we have from~\eqref{ic:eq115} and~\eqref{ic:eq127} that
\begin{align}
\label{ic:eq131}
\frac{b_{k-1}^2}{P_k \beta_{k-1}^2}= B_{k-1}=\frac{\lambda_M^{(k)}-\lambda_M^{(k+1)}}{(A-C)A^{M-k-1}}.
\end{align}
In order for~\eqref{ic:eq131} to have a solution pair $(\beta_{k-1}, P_k>0)$, we need
\begin{align}
\label{ic:eq132}
\frac{\lambda^{(k)}-\lambda^{(k+1)}}{A-C}>0.
\end{align}
Furthermore, from~\eqref{ic:eq114} and~\eqref{ic:eq116}, the following relations must be satisfied
\begin{align}
\label{ic:eq133}
\frac{P_k[1-b_k(1-a)]^2}{P_{k+1}\beta_k^2}=A,
\end{align}
and
\begin{align}
\label{ic:eq134}
\frac{P_k([1-b_k(1-a)]^2+ Mab_k[2(1-a)b_k + Mab_k -2])}{P_{k+1}\beta_k^2}=C.
\end{align}
From~\eqref{ic:eq131} and~\eqref{ic:eq133}, we obtain
\begin{align}
P_k =& \frac{Ab_k^2}{B_k[1-b_k(1-a)]^2},\\
\beta_k^2=&\frac{b_k^2}{P_{k+1} B_k},
\end{align}
for all $k=1,2,3,...,M-1$. Since $P_k>0$ obviously, we need $B_k >0$ and $b_k \notin \{0,1/(1-a)\}$ for all $k$. This condition is satisfied from~\eqref{ic:eq131} and~\eqref{ic:eq132}. 

Besides, we need to set $P_M=P$ and $\lambda_M^{(k)}=\lambda^{(k)}$ for all $k=1,2,...,M$. 

The last thing we need to check is that there exists a $b_k \neq 0$ satisfying~\eqref{ic:eq134}. From~\eqref{ic:eq100},~\eqref{ic:eq102}, and~\eqref{ic:eq133}, this condition is equivalent to that the following equation has at least a non-zero solution $b_k$:
\begin{align}
\label{ic:eq137}
&\frac{P_kMab_k[2(1-a)b_k+Mab_k-2]}{P_{k+1}\beta^2_k}^2 \nn\\
&\qquad =\frac{PMab[2(1-a)b+Mab-2]}{P\beta^2}.
\end{align}
Combing~\eqref{ic:eq137} with~\eqref{ic:eq133}, the requirement becomes
\begin{align}
&\frac{A}{[1-b_k(1-a)]^2} Mab_k[2(1-a)b_k+ Mab_k-2]\nn\\
&\qquad =\frac{PMab[2(1-a)b+Mab-2]}{P\beta^2}.
\end{align}
From~\eqref{ic:eq100}, this relation is satisfied by choosing $b_k=b$ for all $k=1,2,3,...,M-1$.

In short, for the case $A\neq 0, A\neq C, (\lambda^{(k)}-\lambda^{(k+1)})/(A-C)>0, \hspace{1mm} \forall k<M$, \red{we can realize $P_M=P$ and $\lambda_M^{(k)}=\lambda^{(k)}$ for all $k=1,2,\ldots,M$} by setting the parameters as follows:
\begin{align}
b_k&=b,\\
B_k&=\frac{\lambda^{(k+1)}-\lambda^{(k+2)}}{(A-C)A^{M-k-2}}, \hspace{1mm} (\lambda^{(M+1)}:=\lambda^{(1)})\\
\beta_k^2&=\frac{b_k^2}{P_{k+1}B_k},\\
P_k&=\frac{AP_{k+1}\beta_k^2}{[1-b_k(1-a)]^2}.
\end{align}
For $n \geq M$, we only need to set $b_n=b, \beta_n =\beta, P_n=P$ and obtain $\lambda_n^{(k)}=\lambda^{(k)}$ for all $k=1,2,...,M$ from the relation~\eqref{ic:eq66}.

Observe that since $W_n^{(m)}= P$ for all $n\geq M$, we have $W=P < \infty$. Applying Theorem~\ref{ic:thm1} to the above results, Theorem~\ref{ic:thm2} is obtained. Note that, since Theorem~\ref{ic:thm1} holds only for $0< \beta <1$, $\log^+$ must be used instead of $\log$.
\section{Proof of Theorem~\ref{ic:thm3}}\label{proofofthm3}
We will show that we can find a pair $(b,\beta,\lambda)$ such that all the conditions in Theorem~\ref{ic:thm2} are satisfied for each fixed $A>1$.  
From~\eqref{ic:eq99}, we have
\begin{align}
\lambda^{(k+1)}-\lambda^{(k)}=\frac{1}{A}\left(\lambda^{(k)}-\lambda^{(k-1)}\right).
\end{align}
Therefore, the condition $\lambda^{(1)}>\lambda^{(2)} >\cdots > \lambda^{(2)} > \lambda^{(M)}$ is satisfied if $\lambda^{(1)}>\lambda^{(2)}$. This condition is equivalent to
\begin{align}
\lambda > \frac{\lambda-B}{A}. 
\end{align} Of course, this equation is satisfied if we choose $A>1$. 
Moreover, we obtain from~\eqref{ic:eq100} that
\begin{equation}
b=\frac{1\pm \beta \sqrt{A}}{1-a}.
\end{equation}
On the other hand, from~\eqref{ic:eq94},~\eqref{ic:eq100},~\eqref{ic:eq101}, and~\eqref{ic:eq128} we obtain
\begin{align}
\label{ic:eq129}
C=A+\frac{M}{\lambda}(1-A-B).
\end{align}
From~\eqref{ic:eq129}, $A>1$, $B\geq 0$, and $\lambda>0$ we have that $A-C>0$. This means that the condition~\eqref{ic:eq96} is satisfied.

Substituting~\eqref{ic:eq129} into~\eqref{ic:eq97} and noting that $A\neq 1$, we obtain
\begin{align}
\left(1-\left[A+\frac{M}{\lambda}(1-A-B)\right]A^{M-1}\right)\lambda=B\frac{A^M-1}{A-1}.
\end{align}
This equation is equivalent to
\begin{align}
&\left(1-\left[A+\frac{M}{\lambda}(1-A)\right]A^{M-1}\right)\lambda \nn\\
&\qquad=B\left[-MA^{M-1}+\frac{A^M-1}{A-1}\right],
\end{align}
or
\begin{align}
B=\frac{\left(1-\left[A+\frac{M}{\lambda}(1-A)\right]A^{M-1}\right)\lambda}{-MA^{M-1}+\frac{A^M-1}{A-1}}.
\end{align}
Observe that $B=b^2/(P\beta^2)$, so we have
\begin{align}
\label{ic:eq220}
\frac{b^2}{P}= \beta^2 \frac{\left(1-\left[A+\frac{M}{\lambda}(1-A)\right]A^{M-1}\right)\lambda}{-MA^{M-1}+\frac{A^M-1}{A-1}}.
\end{align}
Rewrite~\eqref{ic:eq220} as
\begin{align}
\label{ic:eq221}
\frac{b^2}{P}=\beta^2 \frac{1-A^M}{-MA^{M-1}+\frac{A^M-1}{A-1}}\lambda+\frac{M(A-1)A^{M-1}}{-MA^{M-1}+\frac{A^M-1}{A-1}}\beta^2.
\end{align}
Therefore, from~\eqref{ic:eq94} and~\eqref{ic:eq95} we have
\begin{align}
\label{ic:eq223}
\lambda=\frac{\beta^2(1-A)-\frac{b^2}{P}}{ab[2(1-a)b+Mab-2]}.
\end{align}
Replacing the relation~\eqref{ic:eq223} to~\eqref{ic:eq221}, we obtain
\begin{align}
&\frac{b^2}{P}=\frac{M(A-1)A^{M-1}}{-MA^{M-1}+\frac{A^M-1}{A-1}}\beta^2\nn\\
&\enspace +\beta^2\bigg(\frac{1-A^M}{-MA^{M-1}+\frac{A^M-1}{A-1}}\bigg)\bigg(\frac{\beta^2(1-A)-\frac{b^2}{P}}{ab[2(1-a)b+Mab-2]}\bigg).
\end{align}
Rearranging this relation, we have
\begin{align}
&\frac{b^2}{P} ab[2(1-a)b+Mab-2]\nn\\
&=\frac{(A^M-1)(A-1)}{\left(-MA^{M-1}+\frac{A^M-1}{A-1}\right)}\beta^4  -\frac{b^2(1-A^M)}{P\left(-MA^{M-1}+\frac{A^M-1}{A-1}\right)}\beta^2 \nonumber\\
&\qquad+\frac{M(A-1)A^{M-1}}{\left(-MA^{M-1}+\frac{A^M-1}{A-1}\right)}\beta^2 ab[2(1-a)b+Mab-2].
\end{align}
Then, we have an equation
\begin{align}
\label{ic:eq226}
&a[(M-2)a+2]\frac{b^4}{P}-2 a\frac{b^3}{P} -\frac{(A^M-1)}{P \left(-MA^{M-1}+\frac{A^M-1}{A-1}\right)}\beta^2 b^2 \nn\\
&\qquad -\frac{M(A-1)A^{M-1}}{\left(-MA^{M-1}+\frac{A^M-1}{A-1}\right)}a[(M-2)a+2]\beta^2 b^2 \nonumber\\
&\qquad +\frac{M(A-1)A^{M-1}}{\left(-MA^{M-1}+\frac{A^M-1}{A-1}\right)}2a\beta^2 b \nn\\
&\qquad -\frac{(A^M-1)(A-1)}{\left(-MA^{M-1}+\frac{A^M-1}{A-1}\right)}\beta^4=0.
\end{align}
Using $Y_0$, $Y_1$, $Y_2$ defined by~\eqref{defineY0},~\eqref{defineY1}, and~\eqref{defineY2}, respectively, the equation~\eqref{ic:eq226} can be rewritten as
\begin{align}
\label{ic:eq230}
a[(M-2)a+2]\frac{b^4}{P}-2 a\frac{b^3}{P} + Y_2 \beta^2 b^2 +Y_1\beta^2 b +Y_0 \beta^4=0.
\end{align}
Note that
\begin{align}
b&=\frac{1\pm \sqrt{A}\beta}{1-a},\\
b^2&=\frac{1}{(1-a)^2}(1+A\beta^2\pm 2\sqrt{A}\beta),\\
b^3&=\frac{1}{(1-a)^3}(1\pm (\sqrt{A})^3\beta^3\pm 3\sqrt{A}\beta + 3A\beta^2,\\
b^4&=\frac{1}{(1-a)^4}(1\pm 4\sqrt{A}\beta+ 6A\beta^2\pm 4(\sqrt{A})^3\beta^3+A^2\beta^4).
\end{align}
Substituting these results into the equation~\eqref{ic:eq230}, we attain
\begin{align}
Z_4\beta^4 + Z_3 \beta^3 +Z_2 \beta^2 + Z_1\beta + Z_0=0.
\end{align}
Here, $Z_4$, $Z_2$, $Z_0$ are given by~\eqref{defineZ4},~\eqref{defineZ2},~\eqref{defineZ0}, respectively, and
\begin{align}
Z_3&=\pm 4(\sqrt{A})^3\frac{a[(M-2)a+2]}{P(1-a)^4}\mp \frac{2a}{P(1-a)^3}(\sqrt{A})^3 \nn\\
&\qquad \pm \frac{2Y_2\sqrt{A}}{(1-a)^2}\pm \frac{Y_1\sqrt{A}}{1-a},\\
Z_1&=\pm \frac{a[(M-2)a+2]}{P(1-a)^4}4\sqrt{A}\mp \frac{2a}{P(1-a)^3}3\sqrt{A}.
\end{align}
Last but not least, for $0<\lambda<M$,~\eqref{ic:eq204} must hold. Now, since $A>1$,~\eqref{ic:eq204} holds if and only if
\begin{align}
b[2(1-a)b+Mab-2]<0.
\end{align}
This is equivalent to
\begin{align}
\label{ic:eq242}
0<b<\frac{2}{(M-2)a+2}.
\end{align}
Since we assume that $\beta>0$, it is easy to see that we must choose
\begin{align}
\label{ic:eq243}
b=\frac{1-\beta\sqrt{A}}{1-a}.
\end{align}
Therefore, $Z_3$ and $Z_1$ are given by~\eqref{defineZ3} and~\eqref{defineZ1}, respectively. Finally, from~\eqref{ic:eq243}, it is easy to see that~\eqref{ic:eq242} holds if and only if we choose $\beta$ such that
\begin{align}
\frac{Ma}{[(M-2)a+2]\sqrt{A}} < \beta < \frac{1}{\sqrt{A}}\quad \mbox{for}\quad a<1,\\
\label{ic2018:moo1}
\frac{1}{\sqrt{A}}< \beta < \frac{Ma}{[(M-2)a+2]\sqrt{A}}\quad \mbox{for}\quad a>1.
\end{align}
This means that there exists a triplet $(b,\beta,\lambda)$ which satisfies~\eqref{ic:eq203} and ~\eqref{ic:eq204} and that two these conditions are sufficient conditions for~\eqref{ic:eq94}--\eqref{ic:eq97} to hold for any $A>1$. By Theorem~\ref{ic:thm2}, we conclude that Theorem~\ref{ic:thm3} also holds.
\section{Proof of Corrolary~\ref{ic:cor4}}\label{proofofcor4}
Let $A=P^v$ for some $v>0$ which will be determined later. On the other hand, since 
\begin{align}
\alpha=\frac{\log INR}{\log SNR}=\frac{\log(a^2P)}{\log(P)} 
\end{align}
then $a^2P=P^{\alpha}$ and $a=P^{(\alpha-1)/2}$. For $P$ sufficiently large, by keeping the dominant terms in the nominator and denominator of fractions of polynomials in $P$, we have
\begin{align}
Y_0&= -\frac{(A^M-1)(A-1)}{\left(-MA^{M-1}+\frac{A^M-1}{A-1}\right)} \nonumber \\
&\approx \frac{1}{M-1}P^{2v},\\
Y_1&=\frac{M(A-1)A^{M-1}}{\left(-MA^{M-1}+\frac{A^M-1}{A-1}\right)}2a  \nonumber  \\
&\approx \frac{M P^{vM}}{(-M+1)P^{v(M-1)}}2P^{(\alpha-1)/2}  \nonumber \\
&\approx \frac{-2M}{M-1}P^{(v+\frac{\alpha-1}{2})},\\
Y_2&= -\left(\frac{A^M-1}{P\left(-MA^{M-1}+\frac{A^M-1}{A-1}\right)}\right)\nn\\
&\qquad -\frac{M(A-1)A^{M-1}}{-MA^{M-1}+\frac{A^M-1}{A-1}}a[(M-2)a+2]  \nonumber \\
&\approx \frac{P^{v-1}}{M-1}+\frac{M(M-2)}{M-1}P^{v+\alpha-1}  \nonumber  \\
&\approx \frac{M(M-2)}{M-1}P^{v+\alpha-1}.
\end{align}
Using the same arguments as above, it follows that
\begin{align}
Z_4 &\approx (M-1) P^{2v},\\
Z_3 &\approx - 2M P^{\frac{3v}{2}},\\
Z_2 &\approx \frac{M^2}{(M-1)}P^{v},\\
Z_1 &\approx - (4M-2)P^{\frac{v}{2}-\alpha},\\
Z_0 &\approx M P^{-\alpha}.
\end{align}
The equation~\eqref{ic:eq203} will be satisfied if for $P$ sufficiently large ($P\rightarrow \infty$) we can show that for some $\beta >0$,
\begin{align}
\label{ic:eq266}
&(M-1) P^{2v}\beta^4 - 2M P^{\frac{3v}{2}}\beta^3 \nn\\
&\qquad + \frac{M^2}{M-1}P^{v}\beta^2 -(4M-2)P^{\frac{v}{2}-\alpha}\beta + MP^{-\alpha}\approx 0.
\end{align}
Moreover, $\beta$ must satisfy~\eqref{ic2018:moo1}, which becomes
\begin{align}
P^{-v/2} < \beta < \frac{Ma}{[(M-2)a+2]}P^{-v/2}\approx \frac{M}{M-2}P^{-v/2}.
\end{align}
From~\cite[Theorem 1]{Mohajer2013a}, we know that $d(\alpha) \leq \alpha/2$ for $\alpha >1$ hence $\beta$ cannot decay faster than $\alpha/4$ as $P$ tends to infinity. To show that our coding scheme can achieve the optimal generalized degree of freedom $\alpha/2$, we set $\beta=P^{-\alpha/4} \gamma$ for some $\gamma>0$ which does not depend on $P$ and $\alpha$ and show that all the conditions in the Theorem~\ref{ic:thm3} are satisfied. With this setting, it follows that
\begin{align}
\frac{\alpha}{2}-2\log_P(\gamma) < v < \frac{\alpha}{2}+2\log_P\left(\frac{M}{M-2}\right)-2\log_P(\gamma).
\end{align}
For $P\rightarrow \infty$, we have $v\rightarrow \alpha/2$. Hence, we must set $v=\alpha/2$. We will show that we can find such a $\gamma$ to satisfy the equation~\eqref{ic:eq266}. Indeed,
\begin{align}
&MP^{-\alpha}+(M-1) P^{2v} P^{-\alpha} \gamma^4 - 2M P^{\frac{3v}{2}}P^{-3\alpha/4}\gamma^3 \nn\\
&\qquad  + \frac{M^2}{M-1}P^{v} P^{-\alpha/2}\gamma^2  -(4M-2)P^{\frac{v}{2}-\alpha}P^{-\alpha/2}\gamma \approx 0.
\end{align}
The above equation is equivalent to for $P$ sufficiently large
\begin{align}
\label{ic:eq271}
&MP^{-\alpha}+ (M-1) \gamma^4 - 2M \gamma^3 \nn\\
&\qquad + \frac{M^2}{M-1}\gamma^2  +(4M-2)P^{-3\alpha/4}\gamma \approx 0.
\end{align}
By choosing $\gamma=M/(M-1)$, the equation~\eqref{ic:eq271} is satisfied. Next, we check that with our choices of $\gamma$ and $v=\alpha/2$, $\beta=P^{-\alpha/4}\gamma$, we have $0\leq \lambda \leq M$ for $P$ sufficiently large.

Indeed, for $P$ sufficiently large, we see that
\begin{align}
b=\frac{1-\beta\sqrt{A}}{1-a} \approx (\gamma-1) P^{(1-\alpha)/2}.
\end{align}
Hence,
\begin{align}
b^2 \approx (\gamma-1)^2 P^{1-\alpha}.
\end{align}
Then, from~\eqref{ic:eq223} we have
\begin{align}
\lambda &= \frac{\beta^2(1-A)-b^2/P}{(M-2)a^2 b^2 + 2ab^2 -2ab}  \nonumber \\
&\approx \frac{P^{-\alpha/2}\gamma^2 (1-P^{\alpha/2})-(\gamma-1)^2 P^{-\alpha}}{(M-2)(\gamma-1)^2 P + 2(\gamma-1)^2 P^{1-\alpha/2}-2(\gamma-1)\sqrt{P}}  \nonumber \\
&\approx 0.
\end{align}
This means that $\lambda \searrow 0$ as $P \nearrow \infty$.

By the result of Theorem~\ref{ic:thm1}, for $P$ sufficiently large, the achievable symmetric rate $R_{\rm{sym}}=R_{\rm{sym}}(SNR,\alpha)$ is approximate to
\begin{align}
R_{\mathrm{sym}}(SNR,\alpha) &\geq \frac{1}{2}\log^+\frac{1}{\beta^2}  \nonumber  \\
&\approx \frac{\alpha}{4} \log P.
\end{align}
Therefore, the generalized degree of freedom of our code satisfies
\begin{align}
d(\alpha)=\lim_{\mathrm{SNR} \to \infty} \frac{R_{\mathrm{sym}}(SNR, \alpha)}{(1/2)\log(SNR)}\geq \frac{\alpha}{2}.
\end{align}
Since we know from~\cite[Theorem 1]{Mohajer2013a} that $d(\alpha) \leq \alpha/2$ we have
\begin{align}
d(\alpha)=\frac{\alpha}{2}.
\end{align}
\section{Proof of Corrolary~\ref{ic:cor5}}\label{proofofcor5}
Similarly to the proof in Appendix~\ref{proofofcor4}, we set $A=P^v$ where $v>0$ will be determined later. Using the approximation arguments, in which we keep the dominant terms in nominators and denominators of fractional expressions, we obtain
\begin{align}
Y_0 \approx& \frac{1}{M-1}P^{2v},\\
Y_1 \approx& -\frac{2M}{M-1}P^{v+(\alpha-1)/2},\\
Y_2\approx& \frac{M(M-2)}{M-1}P^{v+\alpha-1}.
\end{align}
It follows that
\begin{align}
Z_0\approx& 0,\\
Z_1 \approx& -2 P^{(\alpha-1)/2+ v/2 -1},\\
Z_2 \approx& -\frac{2M}{M-1}P^{v+(\alpha-1)/2},\\
Z_3 \approx& \frac{2M}{M-1}P^{3v/2+(\alpha-1)/2},\\
Z_4 \approx& \frac{1}{M-1} P^{2v}.
\end{align}
The equation~\eqref{ic:eq203} will be satisfied for $P$ sufficiently large if
\begin{align}
\label{ic:eq292}
&\frac{1}{M-1}P^{2v}\beta^4 +\frac{2M}{M-1}P^{3v/2+(\alpha-1)/2}\beta^3 \nn\\
&\qquad -\frac{2M}{M-1}P^{v+(\alpha-1)/2}\beta^2 - 2P^{(\alpha-1)/2+v/2-1}\beta \approx 0.
\end{align}
From the paper~\cite{Mohajer2013a}, we know that $d(\alpha) \leq 1- \alpha/2$ for $\alpha <1$ hence $\beta$ cannot decay faster than $\alpha/4-1/2$ as $P$ tends to infinity. To show that our coding scheme can achieve the optimal generalized degree of freedom $1-\alpha/2$, we set $\beta=P^{\alpha/4-1/2}\gamma$ where $\gamma \neq 0$ does not depend on $P$ and $\alpha$ and show all the conditions in the Theorem~\ref{ic:thm3} to be satisfied. With this setting, the equation~\eqref{ic:eq292} becomes
\begin{align}
\label{ic:eq293}
&\frac{1}{M-1}P^{2v+\alpha-2}\gamma^4 +\frac{2M}{M-1}P^{3v/2+ 5\alpha/4-2}\gamma^3 \nn\\
&\qquad-\frac{2M}{M-1}P^{v+\alpha-3/2}\gamma^2 - 2P^{3\alpha/4+v/2-2}\gamma \approx 0.
\end{align}
 We will return \red{to} this equality later by judiciously choosing $v>0$. Now, we need to make $0<\lambda< M$. 

In order to satisfy $\lambda >0$, from previous arguments, we need to set
\begin{align}
\label{ic:eq294}
b=\frac{1-\beta\sqrt{A}}{1-a},
\end{align}
and
\begin{align}
\frac{Ma}{[(M-2)a+2]\sqrt{A}}< \beta < \frac{1}{\sqrt{A}}.
\end{align}
Moreover, from~\eqref{ic:eq294} and the choices of $A$ and $\beta$, we have
\begin{align}
b &\approx 1- P^{v/2+\alpha/4-1/2}\gamma, \\
b^2 &=1 +P^{v+\alpha/2-1}\gamma^2 -2 P^{v/2+\alpha/4-1/2}\gamma  \nonumber  \\
&\approx 1-2 P^{v/2+\alpha/4-1/2}\gamma,
\end{align}
if choosing $v < 1-\alpha/2$.  With the choice of $v < 1-\alpha/2$, it also follows that
\begin{align}
b^2 - b &\approx P^{v+\alpha/2-1}\gamma^2 -P^{v/2+\alpha/4-1/2}\gamma  \nonumber  \\
&\approx -P^{v/2+\alpha/4-1/2}\gamma.
\end{align}
From~\eqref{ic:eq223}, we have
\begin{align}
\lambda &\approx \frac{\gamma^2P^{\alpha/2-1}(1-P^v)-P^{-1}(1-2P^{v/2+\alpha/4-1})\gamma}{-2P^{(\alpha-1)/2}P^{v/2+\alpha/4-1/2}\gamma}  \nonumber \\
&\approx \frac{-\gamma^2 P^{\alpha/2-1+v}}{-2 \gamma P^{v/2+3\alpha/4-1}} \nonumber \\
&=\frac{\gamma}{2}P^{v/2-\alpha/4}.
\end{align}
To make $\lambda$ bounded in $(0,M)$ when $P \rightarrow \infty$ and $\lambda >0$, we should choose $v=\alpha/2$ and $0<\gamma< 2M$ such as $\gamma=M$. 

Now, we check the relation~\eqref{ic:eq293} when setting $v=\alpha/2$. It is easy to see that the left hand side of~\eqref{ic:eq293} will be
\begin{align}
&\frac{1}{M-1}P^{2(\alpha-1)}\gamma^4+\frac{2M}{M-1}P^{2(\alpha-1)}\gamma^3 \nn\\
&\qquad -\frac{2M}{M-1}P^{3/2(\alpha-1)}\gamma^2-2 P^{2(\alpha-1)}\gamma\to 0 
\end{align} as $P \rightarrow \infty$ since $\alpha <1$.

This means that for $P$ sufficiently large and $\alpha<1$, by Theorem~\ref{ic:thm2}, the achievable symmetric rate is approximate to
\begin{align}
R_{\rm{sym}}(SNR,\alpha) &\geq \frac{1}{2}\log^+\frac{1}{\beta^2} \nonumber \\
&\approx \frac{1}{2}\bigg(1-\frac{\alpha}{2}\bigg) \log P.
\end{align}
Therefore, the generalized degree of freedom of our code is greater than or equal to
\begin{align}
d(\alpha)=\lim_{SNR\to  \infty} \frac{R_{\rm{sym}}(SNR, \alpha)}{(1/2)\log(SNR)}\geq 1-\frac{\alpha}{2}.
\end{align}
Since we know from~\cite[Theorem 1]{Mohajer2013a} that $d(\alpha) \leq 1 - \alpha/2$ for $\alpha < 1$, we have
\begin{align}
d(\alpha)=1-\frac{\alpha}{2},\quad  \alpha <1.
\end{align}
\subsection*{Acknowledgements}
The authors would like to thank the associate editor Prof.\ Max Costa and the anonymous reviewers for their excellent and detailed comments that helped to improve the presentation of the paper.
\bibliographystyle{IEEEtran}
\bibliography{IEEEabrv,thesisbib}

\begin{thebibliography}{10}
\providecommand{\url}[1]{#1}
\csname url@samestyle\endcsname
\providecommand{\newblock}{\relax}
\providecommand{\bibinfo}[2]{#2}
\providecommand{\BIBentrySTDinterwordspacing}{\spaceskip=0pt\relax}
\providecommand{\BIBentryALTinterwordstretchfactor}{4}
\providecommand{\BIBentryALTinterwordspacing}{\spaceskip=\fontdimen2\font plus
\BIBentryALTinterwordstretchfactor\fontdimen3\font minus
  \fontdimen4\font\relax}
\providecommand{\BIBforeignlanguage}[2]{{%
\expandafter\ifx\csname l@#1\endcsname\relax
\typeout{** WARNING: IEEEtran.bst: No hyphenation pattern has been}%
\typeout{** loaded for the language `#1'. Using the pattern for}%
\typeout{** the default language instead.}%
\else
\language=\csname l@#1\endcsname
\fi
#2}}
\providecommand{\BIBdecl}{\relax}
\BIBdecl

\bibitem{Ahlswede1974}
R.~Ahlswede, ``The capacity region of a channel with two senders and two
  receivers,'' \emph{Ann. Probability}, vol.~2, no.~5, pp. 805--814, 1974.

\bibitem{Carleial1974}
A.~B. Carleial, ``Interference channels,'' \emph{IEEE Trans. on Inform. Th.},
  vol.~24, no.~1, pp. 60--70, 1978.

\bibitem{Han81}
T.~S. Han and K.~Kobayashi, ``{A new achievable rate region for the
  interference channel},'' \emph{IEEE Trans. on Inform. Th.}, vol.~27, no.~1,
  pp. 49--60, 1981.

\bibitem{Bresler2008}
G.~Bresler and D.~Tse, ``The two‐user gaussian interference channel: a
  deterministic view,'' \emph{European Transactions on Telecommunications},
  vol.~19, no.~4, pp. 333--354, 2008.

\bibitem{Etkin2008}
R.~H. Etkin, D.~N.~C. Tse, and H.~Wang, ``Gaussian interference channel
  capacity to within one bit,'' \emph{IEEE Trans. on Inform. Th.}, vol.~54,
  no.~12, pp. 5534--5562, 2008.

\bibitem{Kramer2002a}
G.~Kramer, ``Feedback strategies for white {G}aussian interference networks,''
  \emph{IEEE Trans. on Inform. Th.}, vol.~48, no.~1, pp. 1423--1438, Jan. 2002.

\bibitem{Jiang2007e}
J.~Jiang, Y.~Xin, and H.~K. Garg, ``Empirical processes and typical
  sequences,'' in \emph{41st Annual Conference on Information Sciences and
  Systems}, Baltimore, Maryland, 2007.

\bibitem{Suh2009e}
C.~Suh and D.~Tse, ``Symmetric feedback capacity of the {Gaussian} interference
  channel to within one bit,'' in \emph{Proc. of Intl. Symp. on Inform. Th.},
  Seoul, Korea, 2009.

\bibitem{Suh2011a}
------, ``Feedback capacity of the {Gaussian} interference channel to within 2
  bits,'' \emph{IEEE Trans. on Inform. Th.}, vol.~57, no.~5, pp. 2667--2685,
  2011.

\bibitem{Tandon2013a}
R.~Tandon, S.~Mohajer, and H.~V. Poor, ``On the symmetric feedback capacity of
  the {K}-user cyclic {Z}-interference channel,'' \emph{IEEE Trans. on Inform.
  Th.}, vol.~59, no.~5, pp. 2713--2733, 2013.

\bibitem{Mohajer2013a}
S.~Mohajer, R.~Tandon, and H.~V. Poor, ``On the feedback capacity of the fully
  connected {K}-user interference channel,'' \emph{IEEE Trans. on Inform. Th.},
  vol.~59, no.~5, pp. 2863--2881, 2013.

\bibitem{Ashraphijuo2014e}
M.~Ashraphijuo, V.~Aggarwal, and X.~Wang, ``On the symmetric {K}-user linear
  deterministic interference channels with limited feedback,'' in \emph{52nd
  Annual Allerton Conference}, 2014, pp. 366--371.

\bibitem{Truong2014e}
L.~V. Truong, ``Posterior matching scheme for {Gaussian} multiple access
  channel with feedback,'' in \emph{Proc. of IEEE Information Theory Workshop},
  Tasmania, Australia, 2014, pp. 476--480.

\bibitem{TruongYamamoto2015e}
L.~V. Truong and H.~Yamamoto, ``On the capacity of symmetric {Gaussian}
  interference channels with feedback,'' in \emph{Proc. of Intl. Symp. on
  Inform. Th.}, Hong Kong, China, 2015, pp. 201--205.

\bibitem{TruongYamamoto17a}
------, ``Posterior matching for {Gaussian} broadcast channels with feedback,''
  \emph{IEICE Transactions on Fundamentals}, vol. E100-A, no.~5, pp.
  1165--1178, May 2017.

\bibitem{ShayevitzF}
O.~Shayevitz and M.~Feder, ``Optimal feedback communication via posterior
  matching,'' \emph{IEEE Trans. on Inform. Th.}, vol.~57, no.~3, pp.
  1186--1222, 2011.

\bibitem{Wicker}
S.~B. Wicker, \emph{Error Control Systems for Digital Communication and
  Storage}.\hskip 1em plus 0.5em minus 0.4em\relax Prentice Hall, Inc., 1995.

\bibitem{Cov06}
T.~M. Cover and J.~A. Thomas, \emph{Elements of Information Theory},
  2nd~ed.\hskip 1em plus 0.5em minus 0.4em\relax Wiley-Interscience, 2006.

\bibitem{Ozarow}
L.~H. Ozarow, ``The capacity of the white {Gaussian multiple} access channel
  with feedback,'' \emph{IEEE Trans. on Inform. Th.}, vol.~30, no.~4, pp.
  623--629, 1984.

\bibitem{Ozarow1984}
L.~Ozarow and S.~Leung, ``An achievable region and outer bound for the
  {Gaussian} broadcast channel with feedback (corresp.),'' \emph{IEEE Trans. on
  Inform. Th.}, vol.~30, no.~4, pp. 667--671, 1984.

\bibitem{Shayevitz2007e}
O.~Shayevitz and M.~Feder, ``Communication with feedback via posterior
  matching,'' in \emph{Proc. of Intl. Symp. on Inform. Th.}, Nice, France,
  2007.

\bibitem{SK66}
J.~Schalkwijk and T.~Kailath, ``A coding scheme for additive noise channels
  with feedback--{Part I}: No bandwith constraint,'' \emph{IEEE Trans. on
  Inform. Th.}, vol.~12, no.~2, pp. 172--182, 1966.

\bibitem{SK66Apr}
J.~Schalkwijk, ``A coding scheme for additive noise channels with
  feedback--{Part II}: Band-limited constraint,'' \emph{IEEE Trans. on Inform.
  Th.}, vol.~12, no.~4, pp. 183--189, 1966.

\bibitem{Kramer2004a}
G.~Kramer, ``Correction to ``feedback strategies for white {Gaussian}
  interference networks", and a capacity theorem for {Gaussian} interference
  channels with feedback,'' \emph{IEEE Trans. on Inform. Th.}, vol.~50, no.~6,
  pp. 1373--1374, Jun. 2004.

\end{thebibliography}
 
\begin{IEEEbiographynophoto}{Lan V.\ Truong} (S'12--M'15) received the B.S.E.\ degree in electronics and telecommunications from the Posts and Telecommunications Institute of Technology (PTIT), Hanoi, Vietnam, in 2003, and the M.S.E.\ degree from the School of Electrical and Computer Engineering, Purdue University, West Lafayette, IN, USA, in 2011, and the Ph.D.\ degree from the Department of Electrical and Computer Engineering, National University of Singapore (NUS), Singapore, in 2018. He was an Operation and Maintenance Engineer with MobiFone Telecommunications Corporation, Hanoi, for several years. He spent one year as a Research Assistant with the NSF Center for Science of Information and Department of Computer Science, Purdue University, in 2012. From 2013 to 2015, he was an Academic Lecturer with the Department of Information Technology Specialization, FPT University, Hanoi, Vietnam. Since 2018, he has been a Research Assistant/Post-Doctoral Research Fellow with the Department of Computer Science, School of Computing, NUS. His research interests include information theory, machine learning, and communications.  
\end{IEEEbiographynophoto}

\begin{IEEEbiographynophoto}{Hirosuke Yamamoto} (S'77--M'80--SM'03--F'11--LF'18) was born in Wakayama, Japan, in 1952. He received the B.E. degree from Shizuoka University, Shizuoka, Japan, in 1975 and the M.E. and Ph.D. degrees from the University of Tokyo, Tokyo, Japan, in 1977 and 1980, respectively, all in electrical engineering. In 1980, he joined Tokushima University. He was an Associate Professor at Tokushima University from 1983 to 1987, the University of Electro-Communications from 1987 to 1993, the University of Tokyo from 1993 to 1999, and a Professor at the University of Tokyo from 1999 to 2018. Since 2018, he has been a Professor Emeritus at the University of Tokyo and an Institute Professor at Research and Development Institute, Chuo University. In 1989-1990, he was a Visiting Scholar at the Information Systems Laboratory, Stanford University, Stanford, CA. His research interests are in Shannon theory, data compression algorithms, error correcting codes, and information theoretic cryptology.

Dr. Yamamoto served as the Chair of IEEE Information Theory Society Japan Chapter in 2002-2003, the TPC Co-Chair of the ISITA2004, the TPC Chair of the ISITA2008, the President of the SITA (Society of Information Theory and its Applications) in 2008-2009, the President of the ESS (Engineering Sciences Society) of IEICE in 2012-2013, an Auditor of the IEICE in 2016--2017, an Associate Editor for Shannon Theory, the IEEE TRANSACTIONS ON INFORMATION THEORY in 2007--2010, the Editor-in-Chief for the IEICE Transactions on Fundamentals of Electronics, Communications and Computer Sciences in 2009--2011. He is a Fellow of the IEICE.
\end{IEEEbiographynophoto}
 \end{document}